\documentclass[twocolumn,preprintnumbers,amsmath,amssymb,nofootinbib
,superscriptaddress]{revtex4}
\usepackage{hyperref}
\usepackage{graphicx}
\usepackage{color}
\usepackage{xcolor}
\usepackage{comment}
\usepackage{lmodern}

\newcommand{\nonumbo}{\nonumber \\ && }
\newcommand{\backo}{\!\!\!\!\!\!\!\!\!\!}

\newcommand{\smebo}{\eea }

\newcommand{\smbbo}{\bea  && }
\newcommand{\bl}{\biggl(}
\newcommand{\br}{\biggr)}
\newcommand{\Tr}{\makebox{Tr}}
\newcommand{\vvx}{\vec{x}}
\newcommand{\vvy}{\vec{y}}
\newcommand{\vvr}{\vec{r}}

\newcommand{\vvq}{\vec{q}}

\newcommand{\be}{\begin{equation}}  
\newcommand{\ee}{\end{equation}}  
\newcommand{\bear}{\begin{eqnarray}}   
\newcommand{\eear}{\end{eqnarray}}  
\newcommand{\ba}{\begin{array}}  
\newcommand{\ea}{\end{array}}

\newcommand{\VEV}[1]{\langle #1 \rangle}

\usepackage{diagbox}

\definecolor{rossoCP3}{cmyk}{0,.88,.77,.40}
\definecolor{blueRef}{rgb}{0.2,0.2,0.6}
\definecolor{blue}{rgb}{0,0.396,0.741}
\hypersetup{
	colorlinks, 
	bookmarksopen, 
	bookmarksnumbered,
	citecolor=blueRef, 		
	linkcolor=rossoCP3,	
	urlcolor=rossoCP3,			
}

\newskip\humongous \humongous=0pt plus 1000pt minus 1000pt

\newif\ifdtup

\def\oldreffmt#1{\rlap{[#1]} \hbox to 2\parindent{}}

\def\figfmt#1{\rlap{Figure {#1}} \hbox to 1in{}}  
  
\def\etal{\hbox{\it et al.}}  
  
\def\tr{\mathop{\rm tr}}  
\def\Tr{\mathop{\rm Tr}}

\def\slash#1{#1\!\!\!/\!\,\,} 	
\def\beq{\begin{equation}}  
\def\eeq{\end{equation}}  
\def\bea{\begin{eqnarray}}  
\def\eea{\end{eqnarray}}  
\def\half{\frac{1}{2}}  
  
\def\bq{\begin{quote}}  
\def\eq{\end{quote}}

\def\half{\frac{1}{2}}    
\def \lta {\mathrel{\vcenter  
     {\hbox{$<$}\nointerlineskip\hbox{$\sim$}}}}  
\def \gta {\mathrel{\vcenter  
     {\hbox{$>$}\nointerlineskip\hbox{$\sim$}}}}   

\def \etal {{\it et al.}\ }  
\relax

\newdimen\tdim  
\tdim=\unitlength  
\def\bar{\overline}

\begin{document}
\preprint{FERMILAB-CONF-24-0010-T}

\title{Nambu and Compositeness\footnote{Invited talk, Miami 2003,Topical Conference on Elementary Particles, 
Astrophysics and Cosmology, organizer: Thom Curtright, University of Miami.}
}

\author{Christopher T. Hill}
\email{hill@fnal.gov}
\affiliation{Fermi National Accelerator Laboratory,
P. O. Box 500, Batavia, IL 60510, USA}
\affiliation{Department of Physics, University of Wisconsin-Madison, Madison, WI, 53706
}

\begin{abstract}
\vspace{0.1 in}
The Nambu--Jona-Lasinio model (NJL) \cite{NJL} is the simplest field
theory of a composite scalar boson,
consisting of a pair of chiral fermions. A bound state emerges 
from an assumed point-like 4-fermion interaction and
is described by local effective field, $\Phi(x)$.
We review this in the context of the renormalization group
and show some its phenomenological successes, such as the QCD chiral dynamics and the prediction
of stable heavy-light resonances that reveal the single light quark chiral dynamics.
We also review some NJL--inspired composite Higgs models. 
We then describe a novel UV completion of the NJL model, to an extended interaction,
where the solution is described by a bilocal effective, Yukawa field, $\Phi(x,y)$.
I conclude with a personal recollection of Yoichiro Nambu.
\end{abstract}

\maketitle
 
\date{\today}

\email{hill@fnal.gov}


\section{Introduction}

This is an expanded version of an
invited talk given in an honorary session dedicated to Yoichiro Nambu at 
the Miami 2023 Topical Conference on Elementary Particles, Astrophysics and Cosmology, 
Dec. 13--19,  Fort Lauderdale, Florida.  Based upon
discussions with participants  I was inspired 
to collect things together and present them in this form.  
In particular, I want to emphasize that
the Nambu--Jona-Lasinio model,  NJL \cite{NJL}, often considered a toy model
for chiral dynamics, is a powerful tool and 
has led to bona fide discoveries in physics.

For example: thirty years ago
Bill Bardeen and I applied the NJL model
to heavy--light mesons \cite{BHHQ}. This predicted
a universal mass gap between a ground state, e.g., the $0^-$, 
charm-light $D_{u,d}$ meson, 
$\bar{c}(u,d)$,
and it's ``chiral partner,'' the $0^+$,  $\widetilde{D}_{u,d}$. Hence, these two particles would
 be degenerate if we could turn off chiral symmetry breaking.
We estimated the mass gap in the NJL model
to be $\Delta M \sim 338$ MeV.  This is enough to make $\widetilde{D}_{u,d}$ hopelessly
broad due to the main decay mode $\widetilde{D}_{u,d} \rightarrow D+\pi$, so this seemed 
at first to be an uninteresting study.
However, we realized that the $\bar{c}s$ state, the $\widetilde{D}_s$, would decay as
$\widetilde{D}_s\rightarrow D_{u,d}+K$.  If we do the arithmetic with the estimated 
$\Delta M\sim 338$ MeV we see
the decay mode is kinematically blocked, owing to the large mass of the K meson.
Hence the chiral partner of the $\widetilde{D}_s$, would be a very narrow
resonance and potentially observable!  

Ten years after our original study
the $\widetilde{D}_s(2317)$ was discovered in the BABAR experiment \cite{BABAR}
and the mass gap is measured to be  $\Delta M \sim 349$ MeV, which
agrees with our NJL estimate to a few
percent.  We then joined up with Estia Eichten,  \cite{BEH},
and showed that this is a generic phenomenon and applies to the $(0^+,1^+)$
states, $\bar{b}s$ and $\bar{c}s$, 
and the $\small (\frac{1}{2}^-,\frac{3}{2}^-)$ baryons, 
$[cc]_{J=1}s$, $[bb]_{J=1}s$, $[cb]_{J=1}s$, and $[cb]_{J=0}s$.
So, if you count them all,  we have theoretically ``discovered''
twelve new long-lived resonances, all precisely predicted and observable in experiments.
Remarkably, these provide a unique view of chiral dynamics of single light quarks, and
all of this followed by using the Nambu--Jona-Lasinio model as a guide.

In the present note I review the chiral dynamics of QCD through the NJL model, 
including the heavy-light phenomena.  

I then  go further to discuss the possibility
of composite Higgs bosons.  To me, ascertaining whether the known Higgs boson
is point-like or not is perhaps the most important thing the LHC (or
future colliders, such as an $e^+e^-$ LC or a muon collider) can pursue.

I will also discuss a new effort to extend the NJL model beyond the point-like limit
and to formulate the non-point-like internal dynamics of the bound states it approximates. 
The  NJL model serves as an anchor to the extended dynamics, 
and I will show a surprise emerges here as well.  
I think this may be an important venue for future research.
I conclude with a personal recollection, a story of lunch with Nambu in Hiroshima.  

\section{ Nambu--Jona-Lasinio Model \label{Intro} }

The Nambu--Jona-Lasinio model (NJL) \cite{NJL} is the simplest field
theory of a composite scalar boson,
consisting of chiral fermions. 
An effective {\em point-like} bound state emerges 
from an assumed {\em point-like} 4-fermion interaction.
We begin with a lightning review of the modern solution
of the NJL model using concepts of the renormalization group.

We assume chiral fermions, each with $N_c$ ``colors'' labeled by $(a,b,...)$.
A non-confining,  chiral invariant  $U(1)_{L}\times U(1)_{R}$
action, then takes the form:
\smbbo
\label{NJL1}
S_{NJL}
=\!\int\! d^4x \;\bl i\bar{\psi}^a_L(x)\slash{\partial}\psi_{aL}(x)
+ i\bar{\psi}^a_R(x)\slash{\partial}\psi_{aR}(x)
\nonumbo
\qquad \qquad
+\;
\frac{g^2}{M_0^2}
\bar{\psi}^a_L(x)\psi_{aR}(x)\;\bar{\psi}^b_R(x)\psi_{bL}(x)
\br.
\smebo
This can be readily generalized to a $G_L\times G_R$ chiral symmetry
as in Section III below.

We can rewrite eq.(\ref{NJL1}) in an equivalent form
by introducing the local auxiliary
field $\Phi(x)$:
\smbbo
\label{NJL2}
\backo
S_{NJL}
=\!\int\! d^4x \;\bl i\bar{\psi}^a_L(x)\slash{\partial}\psi_{aL}(x)+ i\bar{\psi}^a_R(x)\slash{\partial}\psi_{aR}(x)
\nonumbo
- M_0^2\Phi^\dagger(x) \Phi(x) + g\bar{\psi}^a_L(x)\psi_{aR}(x)\Phi(x)+h.c.  \br.
\smebo 
Using the $\Phi$ ``equation of motion'' in eq.(\ref{NJL2}) 
reproduces  the 4-fermion interaction of eq.(\ref{NJL1}).

{
\begin{figure}
	\centering
	\includegraphics[width=0.4\textwidth]{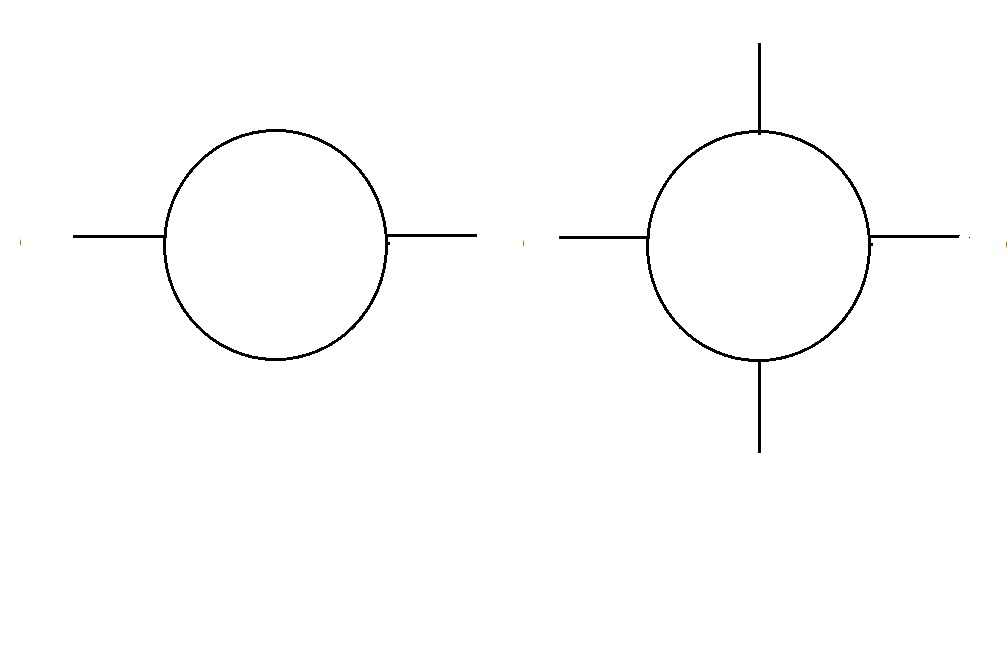}
	\vspace{-0.6in}
	\caption{ Diagrams contributing to the point-like NJL model effective
	Lagrangian, eqs.(\ref{NJL2},\ref{NJL3}). External lines are $\Phi$ and internal
	lines are fermions $\psi$.}
	\label{figvar}
\end{figure}
}

Following Wilson, \cite{Wilson},
we view eq.(\ref{NJL2}) as the action defined at the high energy (short-distance) scale $\mu \sim M$.
We then integrate out the fermions to obtain the effective action for the composite 
field $\Phi$ at a lower scale $\mu<\!\!<M$. 
The calculation in the large-$N_c$ limit
is discussed in detail in \cite{BHL,CTH}.  
The leading $N_c$ fermion loop yields the result
for the $\Phi$ terms in the action at a new scale $\mu$:
\smbbo
\label{NJL3}
\backo
S_{\mu}
=\!\int\! d^4x \;\bl i\bar{\psi}^a_L\slash{\partial}\psi_{aL}+ i\bar{\psi}^a_R\slash{\partial}\psi_{aR}
+
Z\partial_\mu \Phi^\dagger\partial^\mu \Phi
\nonumbo
\backo
-m^2\Phi^\dagger \Phi - \frac{\lambda }{2}(\Phi^\dagger \Phi)^2 + g\bar{\psi}^a_L\psi_{aR}\Phi(x)+h.c. 
\br.
\smebo 
where the diagrams of Fig.(1) yield,
\smbbo
\label{NJL4}
\backo
m^2 = M_0^{2}\!-\!\frac{N_{c}g^{2}}{8\pi ^{2}} ( M_0^{2}-\mu^2),
\nonumbo
\backo
Z=\frac{N_{c}g^{2}}{8\pi ^{2}}\ln(M_0/\mu), \;\;\;
\lambda=\frac{N_{c}g^{4}}{4\pi ^{2}}\ln( M_0/\mu).
 \smebo
Here $M_0^2$ is the UV loop momentum 
cut-off, and we include the induced kinetic and quartic interaction terms.
The one-loop result can be improved by using the full renormalization group (RG) \cite{BHL,CTH}.
Hence the NJL model is driven by fermion loops, which are $\propto \hbar$ intrinsically
quantum effects.

Note the behavior of the composite scalar boson mass, 
$m^2$, of eq.(\ref{NJL4}) in the UV.
The $ -{N_{c}g^{2}M_0^{2}}/{8\pi ^{2}}$  term arises from the negative
quadratic divergence in the loop involving the
pair $\left( {\psi }_{R},\psi _{L}\right) $ of Fig.(1), with
 UV cut-off  $M_0^{2}$. 
Therefore, the NJL model has a critical
value of its coupling defined by the cancellation of the large $M_0^2$ terms,
\smbbo
g_{c}^{2}=\frac{8\pi ^{2}}{N_{c}}+{\cal{O}}\bl\!\frac{\mu^2}{M_0^2}\!\br.
\smebo
Note that $\mu$ is the running RG mass, and comes from the lower limit
of the loop integrals and can in principle be small, 
$\mu^2< \!\!<m^2$. 

We can renormalize, $\Phi\rightarrow \sqrt{Z}^{-1}\Phi$, hence:
\smbbo
\label{NJL5}
\backo
S_{\mu}
=\!\int\! d^4x \;\bl i\bar{\psi}^a_L\slash{\partial}\psi_{aL}+ i\bar{\psi}^a_R\slash{\partial}\psi_{aR}
+
\partial_\mu \Phi^\dagger\partial^\mu \Phi
\nonumbo
\backo
-m_r^2\Phi^\dagger \Phi - \frac{\lambda_r }{2}(\Phi^\dagger \Phi)^2 + g_r\bar{\psi}^a_L\psi_{aR}\Phi(x)+h.c. 
\br.
\smebo 
where,
\smbbo
\label{NJL30}
\backo
m_r^2 = \frac{1}{Z}\bl M_0^{2}\!-\!\frac{N_{c}g^{2}}{8\pi ^{2}} ( M_0^{2}-\mu^2)\br
\nonumbo
\backo
g^2_r \!=\! \frac{g^2}{Z}\!=\!\frac{8\pi^2}{N_c\ln(M_0/\mu)},\;\;
\lambda_r\!=\!\frac{\lambda}{Z^2}\! =\!\frac{16\pi ^{2}}{N_c\ln( M_0/\mu)}.
 \smebo

For  super-critical coupling, $g^2>g^2_c$,
we see that $m_r^2<0$ and there will be a vacuum instability.
The effective action, with the induced quartic  $\sim \lambda_r(\Phi^\dagger\Phi)^2$ term, 
is then the usual sombrero potential.
The chiral symmetry is spontaneously broken and the field $\Phi$ acquires a VEV,
\smbbo
\langle\Phi\rangle =v = \frac{m_r}{\sqrt{\lambda_r}},
\smebo
and the
chiral fermions acquire mass, 
\smbbo
m_f=g_r v
\smebo
The physical radial mode (``Higgs'' boson), defined as $ \sqrt{2} Re(\Phi)$, has a mass $m_h$
given by,
\smbbo 
m_h^2=2\lambda v^2.
\smebo 
while the Nambu-Goldstone mode, $Im(\Phi)$, is massless.
Hence we see that the NJL model yields a prediction
for the radial mode,
\smbbo
m_h = \frac{\sqrt{2\lambda_r}}{g_r}m_f=2m_f .
\smebo
The predicted mass, 
$m_h = 2m_f$,
may seem like a free field relation, but in fact the binding involves the fermion
loops that extend  from $\mu$ to $M_0$, so the composite Higgs boson is tightly bound
and effectively point-like up to the scale $M_0$.
The 
Nambu-Goldstone bosons also emerge
as point-like bound states.

Fine-tuning of $g^2 \rightarrow g_c^2$ is possible 
if we want a theory with a hierarchy, $|m_r^2|<\!\!<M_0^2$.
In that case we can appeal to the behavior of the renormalized 
couplings as $\mu\rightarrow M_0$. We see then that both $g_r$
and $\lambda_r$ diverge in the ratio $\lambda_r/g^2_r\rightarrow 2$.
This can be used as a boundary condition on the full
RG evolution of $g_r$ and $\lambda_r$
including gauge fields and scalar interactions themselves.

When applied to a Higgs boson composed of top and anti-top 
this yields $m_r\sim m_{top}\sim 220$ GeV (the quasi-infrared fixed point \cite{PR}\cite{FP}), 
and $m_{Higgs}\sim 260$ GeV \cite{BHL}.
While the top mass prediction isn't too far off (about $20$\% too high), the Higgs mass is significantly
larger than the observed $125$ GeV. It's possible to get the top
mass concordant with the observed value of $175$ GeV by extending the
model to include more Higgs doublets, but
the Higgs mass is difficult to accommodate in the NJL model approximation.
More promising may be a seesaw with a second sequential  
composite Higgs doublet composed of the b-quark (see Section IV.)

\section{Application to QCD Chiral Dynamics}

\subsection{Low mass states}

The NJL model can be used to understand the chiral dynamics of QCD 
\cite{BHHQ}\cite{Bijnens}\cite{Bardeen:1993ae}, and is predictive.
The effective Lagrangian in the light quark sector is
approximated by:
\beq
\label{lq}
{\cal{L}} = \overline{\psi}(i\slash{\partial} - m_q )\psi
-\frac{g^2}{\Lambda^2}\overline{\psi}
\gamma_\mu\frac{\lambda^A}{2}\psi \;\overline{\psi}
\gamma^\mu\frac{\lambda^A}{2}\psi
\eeq
where $\psi$ is an $N_{c}=3$  color triplet and $\lambda^A$ are $SU(3)$ color matrices.

 The interaction in eq.(\ref{lq}) can be viewed as a single gluon exchange potential,
where we assume a fake
``gluon mass'' $\Lambda/\sqrt{2}$, where $\Lambda\sim 1$ GeV. We integrate out
the massive gluon and  truncate on dim $=6$
operators, yielding the effective Lagrangian at  $\mu^2 \sim \Lambda^2$. 
For concreteness we will take a flavor isodoublet,
$\psi = (u, d)$, and in
the limit that the quark mass matrix $m_q\rightarrow 0$, we
have an exact chiral $SU(2)\times SU(2)\times U(1)$ invariant
Lagrangian.

Upon Fierz--rearrangement of the interaction Lagrangian,
keeping only leading terms in
$1/N_{c}$,  eq.(\ref{lq}) takes the form:
\smbbo
\backo
\label{lq2}
 {\cal{L}}_L  =
\overline{\psi}(i\slash{\partial} - m_q )\psi \nonumbo 
+
\frac{g^2}{\Lambda^2}\left(
\overline{\psi}_L\psi_{R}\overline{\psi}_{R}\psi_{L} +
\overline{\psi}_L\tau^A\psi_{R}\overline{\psi}_{R}\tau^A\psi_{L}\right)
\smebo
where $\psi_L = (1-\gamma_5)\psi/2$,
$\psi_R = (1+\gamma_5)\psi/2$ and $\tau^A$ are isospin matrices acting.
Here we have truncated eq.(\ref{lq2}) on the pure
NJL terms which represent the most attractive channels.\footnote{Here we
have dropped the the (vector)$^2$ and
(axial-vector)$^2$ terms which are also generated,
but play no significant role in the chiral dynamics. These
are associated with the formation of virtual $\rho$ and
$A_1$ vector mesons in the model.}

We can solve the light--quark dynamics in large--$N_{c}$
by 
introducing the local auxiliary
field $\Sigma(x)$,
 as we did with $\Phi(x)$ in eq.(\ref{NJL2}), thus yielding an
equivalent effective Lagrangian of the form:
\smbbo
\backo
{\cal{L}} =  \overline{\psi}(i\slash{\partial} - m_q )\psi
- g\overline{\psi}_L\Sigma\psi_R - g\overline{\psi}_R\Sigma^\dagger\psi_L
\nonumbo
- \half\Lambda^2\Tr(\Sigma^\dagger\Sigma)
\smebo
where:
\smbbo
 \Sigma = \half\sigma I_2 + i\pi^a\frac{\tau^a}{2}
\smebo
  $\Sigma$
is a $2\times 2$
complex field, i.e., both
$\sigma$ and $\pi^a$ are complex.\footnote{Otherwise, with $\sigma$ and $\pi$ real
there would be
unwanted contributions from $\langle{T \Sigma\;\Sigma}\rangle
= \langle{T \Sigma^\dagger \;\Sigma^\dagger }\rangle \neq 0$ in integrating
out $\Sigma$.} Hence, there is {\em parity doubling} at this stage, and note that
$Im(\sigma)$ is the $\eta$ meson in the $SU(2)\times SU(2)\times U(1)$ model (it
would become the $\eta'$ in an $SU(3)\times SU(3)\times U(1)$
generalization).

We now integrate out the fermion fields
on scales $\mu^2 < \Lambda^2$, keeping only the leading large--$N_{c}$ fermion
loop contributions of Fig.(1), treating
$\Sigma$ as a classical background field.
Thus we arrive at an effective field theory at the
scale $\mu$:
\bea
{\cal{L}} & = & \overline{\psi}(i\slash{\partial} - m_q )\psi
- g\overline{\psi}_L\Sigma\psi_R - g\overline{\psi}_R\Sigma^\dagger\psi_L
\nonumber \\
& & + Z_2 \Tr(\partial_\mu\Sigma^\dagger\partial^\mu\Sigma) - V(\Sigma)
\eea
where:
\smbbo
\backo
Z_2 = \frac{g^2N}{16\pi^2}\ln(\Lambda^2/\mu^2),\;\;\;\;\lambda=\frac{g^4N}{16\pi^2}\ln(\Lambda^2/\mu^2),
\smebo
and,
\smbbo
\backo
V(\Sigma) =  \bl\half\Lambda^2 - \frac{g^2N}{8\pi^2}(\Lambda^2 - \mu^2)
\br\Tr(\Sigma^\dagger\Sigma) 
\nonumbo
\backo\!\! -
 \frac{gN}{8\pi^2}(\Lambda^2 - \mu^2)\Tr (m_q\Sigma + h.c.)
 +  \lambda\Tr(\Sigma^\dagger\Sigma
\Sigma^\dagger\Sigma).
\smebo
We see that $Z_2\rightarrow 0$ as $\mu\rightarrow \Lambda$,
reflecting the compositeness of the $\Sigma$ field.
Let us now renormalize the $\Sigma$ field:
\beq
\Sigma\rightarrow (Z_2)^{-1/2}\Sigma,
\eeq
and we have the effective Lagrangian at
the scale $\mu$:
\smbbo
{\cal{L}}  =  \overline{\psi}(i\slash{\partial} - m_q )\psi
- \widetilde{g}\overline{\psi}_L\Sigma\psi_R -
\widetilde{g}\overline{\psi}_R\Sigma^\dagger\psi_L
\nonumbo \qquad
+ \Tr(\partial_\mu\Sigma^\dagger\partial^\mu\Sigma) - \widetilde{V}(\Sigma),
\smebo
where:
\smbbo
\backo
\widetilde{g} \equiv  1/\sqrt{Z_2}
\nonumbo
\backo
m^2_\sigma = 
\left(\frac{1}{Z_2}\right)
\left[\half \Lambda^2 - \frac{g^2N_c}{8\pi^2}(\Lambda^2 - \mu^2)
\right]
\nonumbo
\backo
\omega  =   \frac{\widetilde{g}gN_c}{8\pi^2  }(\Lambda^2 - \mu^2),
\;\;\; \widetilde\lambda=\widetilde{g}^2=\frac{16\pi^2}{N_c\ln(\Lambda^2/\mu^2)},
\smebo
and,
\bea
\backo
\widetilde{V}(\Sigma) &= & m_\sigma^2\Tr(\Sigma^\dagger\Sigma) -
\omega\Tr (m_q\Sigma + h.c.)
\nonumbo  +  \widetilde\lambda \Tr(\Sigma^\dagger\Sigma
\Sigma^\dagger\Sigma).
\eea
The effective Lagrangian is seen to be a linear $\sigma$--model
at scales $\mu < \Lambda$.
The theory will develop a chiral instability,
hence a constituent quark mass, provided that $m_\sigma^2$ becomes
tachyonic (negative) at some scale $\mu_0$.
By the choice of the bare coupling constant, $g^2$, we can put the
model in a symmetric phase, $m_\sigma^2>0$ where ${g^2N_c}/{4\pi^2}<1$,
or in a chiral symmetry breaking phase: $m_\sigma^2<0$ where ${g^2N_c}/{4\pi^2}>1$,
where the critical bare coupling corresponds
to $m_\sigma^2=0 $ as $\mu_0\rightarrow 0$.

In the broken phase (ignoring $m_q$)
the $\sigma$ field develops a vacuum
expectation value, $\langle{\sigma}\rangle = f_\pi =
\sqrt{2}|m_\sigma |/\sqrt{\lambda}$, given by:
\beq
\VEV{\sigma}  = \frac{Z_2\Lambda^2}{\lambda}  
\left( \frac{g^2N_c}{4\pi^2} - 1\right)
= \left(\frac{\Lambda^2}{g^2 }
\right)
\left( \frac{g^2N_c}{4\pi^2} - 1\right)
\eeq
 In the broken phase we can then write
$\sigma= f_\pi + \hat{\sigma}$, and
the physical mass$^2$ of the $\hat{\sigma}$ is readily
seen to be $m_{\hat{\sigma}}^2 =2 |m^2_\sigma|$, while the fermion mass becomes
$m_0=\half f_\pi \widetilde{g}$.  Thus, using  $\widetilde{g}^2=\lambda$,
we obtain  the usual
Nambu--Jona--Lasinio result: $m_{\hat{\sigma}} = 2m_0$.

Note that the induced interaction term
$\sim \Tr(\Sigma^\dagger \Sigma \Sigma^\dagger \Sigma)$,
has lifted the degeneracy of the $Re(\pi)$ and $Im(\pi)$ fields.
The $Im(\pi)$ states have become heavy resonances and the $Re(\pi)$ states will 
emerge as Nambu-Goldstone bosons
dynamically at low energies.\footnote{We must go beyond the NJL model
and add a nonperturbative $e^{i\theta}\det(\Sigma) + h.c.$ term to 
lift the low mass $Im(\sigma)$ mode in an $SU(3)\times SU(3)$
theory, which is the heavy $\eta'$.}
The solution to the theory can then be written as a chiral quark model
in which we have both constituent quarks described by $\psi$ and
the mesons described by $\Sigma$. 

In the broken phase
it is useful to pass to a nonlinear $\sigma$--model and write:
\beq
\Sigma \rightarrow \half f_\pi \exp(i\pi^a\tau^a/f_\pi),
\eeq
and:
\smbbo
{\cal{L}}  = \overline{\psi}(i\slash{\partial} - m_q )\psi
- m_0\overline{\psi}_L\exp(i\pi^a\tau^a/f_\pi)\psi_R
\nonumbo
- m_0
\overline{\psi}_R \exp(-i\pi^a\tau^a/f_\pi) \psi_L
\nonumbo+ \Tr(\partial_\mu\Sigma^\dagger\partial^\mu\Sigma)
+ \omega\Tr (m_q\Sigma + h.c.),
\smebo
where $m_0=\half \widetilde{g}f_\pi$ is the constituent quark mass.
Note, in our present normalization conventions that $f_\pi =
93$ MeV. 
In the broken phase we can replace $\sigma = f_\pi$ and $\Sigma
\rightarrow \half f_\pi \exp(i\pi^a\tau^a/f_\pi)$.
We can often replace
${g}\widetilde{\sigma}/2 = {g}\sigma/2 + m_q\sqrt{Z_2}$
since it easy to restore the explicit chiral symmetry breaking quark mass
terms.

By a chiral redefinition of the fields,
$\psi_R \rightarrow \xi\psi_R$ and $\psi_L\rightarrow \xi^\dagger\psi_L$,
where $\Sigma\equiv\xi^{2}$,
we arrive at the Georgi--Manohar Lagrangian (\cite{Manohar}, e.g., their eq.(2.9)).

\subsection{ Heavy-Light Mesons}

Heavy-light (HL) systems involve a single valence light--quark,  and a heavy-quark (or
a pair of heavy quarks that form a ``diquark nucleus''). These systems can be viewed as a 
constituent light quark ``tethered" to the inertially fixed heavy quark.
In QCD the light--quark chiral symmetry is
spontaneously broken as described via chiral Lagrangians with nearly massless pions.

The dynamics can be treated in an NJL model that contains both heavy quarks
with heavy quark symmetry \cite{HQ}\cite{Georgi1}\cite{Wise}, together with light quark chiral symmetry
as in the previous section, and the four fermion interaction inspired by QCD \cite{BHHQ}.
We again solve the theory by
factorizing the heavy--light interaction into auxiliary
background fields with Yukawa couplings
to heavy and light quark vertices.
Upon integrating out the quarks on scales  $\Lambda$
to $\mu $, the auxiliary fields acquire induced kinetic terms
and thus become dynamical heavy--light mesons, in parallel
to the previous examples.

The essential feature of the dynamical chiral symmetry 
breaking is a
mass--splitting in heavy-light fields, between normal parity $(0^-,1^-)$ (s-wave)  ground states
and abnormal parity, $(0^+,1^+)$, (p-wave) partner states. Heavy quark symmetry implies that spin-0 and spin-1 
states of a given parity are approximately degenerate, 
and thus form multiplets.
The key idea is that  a $|s^+\rangle+|s^-\rangle $
is a ``right-handed'' chiral state, transforming as
$(0,3)$ under $SU(3)_R$, while $|s^+\rangle-|s^-\rangle $ is ``left-handed,'' $(3,0)$ under $SU(3)_L$,
where spin $s=(0,1)$.
The mass splitting is then caused by the light quark $(\bar{3}, 3)$ condensate $\langle\sigma\rangle
\sim f_\pi$,
discussed in section III.A.

This generally leads to broad resonances for the heavier $(0^+,1^+)$ states
in $c\bar{u}$ and $c\bar{d}$ mesons as they transition to the lighter $(0^-,1^-)$
groundstates, emitting a pion.
But in the case of  heavy-strange systems, the heavier $c\bar{s}$  $(0^+,1^+)$ states
would have a principle
decay mode to the lighter $(0^-,1^-)$ non-strange
groundstates,  $c\bar{s}\rightarrow c(\bar{u},\bar{d})+ K $.
{\em This is kinematically blocked by the large kaon mass, }
hence the  $c\bar{s}$, $(0^+,1^+)$ states becomes  a set of
narrow, observable resonances. 

This prediction 
was subsequently confirmed by the BABAR experiment  \cite{BABAR}, and 
the predicted mass splitting from the NJL model
agrees to within a few percent with the experimental observation.  Taken together
this predicts four narrow $(0^+,1^+)$ heavy-strange resonances in the $cs$, $bs$.
It also applies to the eight heavy-heavy-strange $(\frac{1}{2}^-,\frac{3}{2}^-)$ baryons, 
$[cc]_{J=1}s$, $[bb]_{J=1}s$, $[cb]_{J=1}s$, and $[cb]_{J=0}s$
 \cite{BEH}.\footnote{
A heavy $[QQ]$ pair, which has color $3\times 3=\bar{3}$, behaves like a single heavy $\bar{Q}$.}

A heavy--light fermion sector can be written as an NJL model extending the interaction
of eq.(\ref{lq}) to include a heavy quark $Q$:
\smbbo
\backo
\label{HNJL}
{\cal{L}}_{HL} =  \overline{Q}(i\slash{\partial} - M) Q
-\frac{2g^2}{\Lambda^2}\overline{Q}
\gamma_\mu\frac{\lambda^A}{2}Q \; \overline{\psi}
\gamma^\mu\frac{\lambda^A}{2}\psi.
\smebo
Here we may generally take $Q=(b,c..)$, $M$ to be the heavy quark mass, and  
$g$ 
the effective coupling at the scale $\Lambda$.
Upon Fierz--rearrangement of the interaction,
(again keeping only leading terms in
$1/N_{c}$ and writing in terms of color singlet densities and isospin indices, $i$),
eq.(\ref{HNJL}) takes the form:
\bea
\label{HINT}
{\cal{L}}_{HL} \!\!& = &\!\! \overline{Q}(i\slash{\partial} - M) Q
+ \frac{g^2}{\Lambda^2}\left(
\overline{Q}\psi_{i}\overline{\psi}^iQ -
\overline{Q}\gamma^5\psi_{i}\overline{\psi}^i\gamma^5 Q \;\right.
\nonumber \\
& &
\left. -\frac{1}{2}\overline{Q}\gamma_\mu\psi_i\overline{\psi}^i\gamma^\mu
Q
-\frac{1}{2}\overline{Q}\gamma_\mu\gamma_5 \psi_i
\overline{\psi}^i\gamma_\mu\gamma_5 Q \right)
\eea

Following Georgi, we introduce a projection operator on the heavy quark field
with a well defined, and approximately conserved, four--velocity $v_\mu$, \cite{Georgi1},
\smbbo
P^\pm = \frac{1\pm\slash{v}}{2}
\nonumbo
Q_v = P^+ \exp(-iMv\cdot x)Q(x)
\smebo
The HQ kinetic term then takes the form:
\beq
\overline{Q}_v iv^\mu{\partial}_\mu Q_{v}
\eeq
 and
eq.(\ref{HINT}) becomes:
\smbbo
\label{HSPIN}
\backo\;\;
{\cal{L}}_{HL} = \overline{Q}_v iv^\mu{\partial}_\mu Q_{v}
+
\frac{g^2}{2\Lambda^2}\bl
\overline{Q}_v\psi_{i}\overline{\psi}^iQ_v -
\overline{Q}_v\gamma^5\psi_{i}\overline{\psi}^i\gamma^5 Q_v \; 
\nonumbo \backo\;\;
 - \overline{Q}_v\gamma_\mu P^-\!\psi_i
\overline{\psi}^iP^-\!\gamma^\mu Q_v
+\overline{Q}_v\gamma_\mu P^-\!\gamma_5 \psi_i
\overline{\psi}^i\gamma_5P^-\!\gamma_\mu Q_v \br
\smebo
The interaction  can be factorized by introducing
heavy auxiliary fields, $(B,B')$ which represent
heavy-light bound states of four velocity $v_\mu$.
To do so {\em we must
introduce four independent fields:} $B\; ( B^5)$ are $0^+$
($0^-$) scalars, while $B_\mu\; (B_\mu^5)$ are
$1^-$ ($1^+$) vectors. 
We
assemble the auxiliary fields into
complex $\bf 4 $ multiplets under
$O(4)=SU(2)_h\times SU(2)_l$, where $SU(2)_h$ ($SU(2)_l$) is the
little group of rotations
which preserves $v_\mu$.
One  spin {\bf 4} multiplet
consists  of the $0^+$ scalar together with
the abnormal parity ($1^+$) vector as $(B, B^{5\mu})$
(isospin indices are understood):
\bea
{\cal{B}}'  =  (i\gamma^5 B + \gamma_\mu
B^{5\mu})P^+
\eea
The other $\bf 4 $ multiplet
consists of the usual $0^-$ scalar and a $1^-$ vector $(B^5, B^{\mu})$:
\bea
{\cal{B}}   =  (i\gamma^5 B^5 + \gamma_\mu
B^{\mu})P^+
\eea
Under heavy spin $O(4)=SU(2)_h\times SU(2)_l$
rotations the $(B, B^{5\mu})$ mix analogously
to $(B^5, B^\mu)$. Note that $v_\mu B^\mu = 0$ always.

 The factorized heavy--light
interaction Lagrangian then takes the compact form:
\bea
\label{H12}
{\cal{L}}_{HL}  & = &\overline{Q}_v iv^\mu{\partial}_\mu Q_{v}
+{g}\overline{Q}_{v}\;(-i\overline{\cal{B}'}^i  \gamma^5
+ \overline{\cal{B}}^i)\;\psi_{i} + h.c.
\nonumber \\
& & +
\Lambda^2 \left[\Tr({\cal\overline{B}B}) + \Tr({\cal\overline{B}'{B}'})
\right]
\eea
Eq.(\ref{H12})
is exactly equivalent to the full four--fermion theory eq.(\ref{HINT})
upon integrating out the ${\cal{B}}$ fields.

For weak coupling, $g$, the linear chiral
invariance is realized and the theory must contain
parity doubled meson states. 
The chiral representations of the composite fields
in the effective
theory are the following combinations:
\bea
& & {\cal{B}}_{1} = \frac{1}{\sqrt{2}}\left({\cal{B}} - i {\cal{B}}'\right)
\qquad
{\cal{B}}_{2} = \frac{1}{\sqrt{2}}\left({\cal{B}} + i {\cal{B}}'\right)
\eea
The full effective Lagrangian for the heavy mesons is derived at loop level by
integrating out the heavy and light quarks, $Q$ and $\psi$, in Eq.(\ref{H12}).
The loop integrals involve the heavy quark propagator $\sim i/v\cdot \ell$
and run over momentum
scales $\mu < \ell < \Lambda$, and we keep only the leading large-$N_c$
induced terms. The details of the
explicit calculations are given in \cite{BHHQ}.

With $\Sigma = \half(\sigma + i\pi\cdot \tau)$, which emerges 
upon including the  the light quark
sector (as discussed above in the section III.A),  
performing a conventional wave--function renormalization
and several field redefinitions, we arrive
at the  full effective action valid to O$(\mu/\Lambda)^2$ \cite{BHHQ}:
\bea
\label{misco}
{\cal{L}}_{LH} & = & -i\half \Tr(\overline{{\cal{B}}}_1 v\cdot\partial
{\cal{B}}_{1} )
 -i\half \Tr(\overline{{\cal{B}}}_{2}  v\cdot\partial  {\cal{B}}_{2} )
\nonumber \\
& & -\frac{{g_{r}} }{2}
\left[  \Tr(\overline{{\cal{B}}_{1}}\Sigma^\dagger
 {\cal{B}}_{2}) +\Tr(\overline{{\cal{B}}_{2}}\Sigma
 {\cal{B}}_{1})
\right]
\nonumber \\
&  &
+\frac{ik_r}{ 2f_\pi}
\left[  \Tr(\overline{{\cal{B}}_{1}}\gamma^5(\slash{\partial}\Sigma^\dagger)
 {\cal{B}}_{2}) - \Tr(\overline{{\cal{B}}_{2}}\gamma^5(\slash{\partial}\Sigma)
 {\cal{B}}_{1})
\right]
\nonumber \\
& & +\left(\Delta_{} + \frac{h_{r}}{\Lambda}\Sigma^\dagger \Sigma\right)
\left[  \Tr(\overline{{\cal{B}}_{1}} {\cal{B}}_{1})
+\Tr(\overline{{\cal{B}}_{2}} {\cal{B}}_{2})
\right]
\
\eea
The parameters of this Lagrangian are determined by loop integrals as:
\bea
g_r & = & \frac{g}{\sqrt{Z_2}};
\qquad
h_r  =  \frac{2g^2 \sqrt{Z_2}\Lambda}{Z_1} ;
\qquad
k_r  =  \frac{2g f_\pi \sqrt{Z_2}}{Z_1 };
\nonumber \\
\Delta & = & \frac{1}{Z_1}\left( \Lambda^2 - Z_1(\Lambda + \mu)/2\pi \right),
\eea
where:
\bea
\backo
Z_1 & = &
\frac{g^2N_c}{8\pi }( \Lambda  - \mu);
\qquad
 Z_2  =  \frac{g^2N_c}{16\pi^2}\left[\ln(\Lambda^2/\mu^2)\right].
\eea
Note $Z_1$ is linear in $\Lambda$ and $\mu$.

The parameters  $g_r$, $h_r$ and $k_r$ are dimensionless,
 and are determined by fitting the
observables of the model.  We view
$\mu$ to be of order the light quark constituent mass, and 
henceforth will be neglected in the expression for $Z_1$.  Note that
terms like $\overline{\cal{B}'}\gamma^5
(\Sigma^\dagger\slash{\partial}\Sigma) {\cal{B}}$ are potentially induced,
but they are subleading as $\sim O(1/\ln(\Lambda/\mu)$, relative to
the terms we keep.

Eq.(\ref{misco})) is manifestly  invariant
under $SU(2)_L\times SU(2)_R$ where the fields transform as:
\smbbo
\backo
 {\cal{B}}_{1} = \bl 0, \half\br \qquad {\cal{B}}_{2} = \bl \half,0 \br
\qquad \Sigma  = \bl \half,\half \br
\smebo
In the spontaneously broken chiral symmetry phase
we can pass to the
the nonlinear realization,
\beq
\Sigma = \half f_\pi
\exp(i\pi\cdot\tau/f_\pi)  \qquad \Sigma = \half f_\pi \xi^2
\eeq
and to the ``current form''
by performing the chiral field redefinitions:
\bea
& & {\cal{B}}_{1} \rightarrow \xi^\dagger  {\cal{B}}_{1}\qquad
{\cal{B}}_{2} \rightarrow \xi  {\cal{B}}_{2}
\eea
where the chiral currents are \cite{BHHQ}:
\smbbo
\backo
{\cal{J}}_{\mu, L}  =  i \xi \partial_\mu \xi^\dagger, \qquad
{\cal{J}}_{\mu,R}  = i \xi^\dagger  \partial_\mu \xi\qquad
\nonumbo
\backo
{\cal{A}}_\mu  = \half({\cal{J}}_{\mu,R } - {\cal{J}}_{\mu,L} ),\;\;\;\;
{\cal{V}}_\mu  = \half({\cal{J}}_{\mu,R }  + {\cal{J}}_{\mu,L} ).
\smebo
As usual the ${\cal{J}}_A$ are matrices acting
on the isospin indices of meson fields.
The mass matrix is diagonal 
in the current form by:
\beq
\widetilde{\cal{B}} = \frac{1}{\sqrt{2}} ({\cal{B}}_1 + {\cal{B}}_2),
\qquad
\widetilde{\cal{B}}' = \frac{i}{\sqrt{2}} ({\cal{B}}_1 - {\cal{B}}_2).
\eeq
The Lagrangian then becomes:
\bea
{\cal{L}}_{LH} & = & -\half \Tr(\overline{\widetilde{\cal{B}}}
v\cdot (i{\partial}
+ {{\cal{V}}}){\widetilde{\cal{B}}} )
 -\half \Tr(\overline{\widetilde{\cal{B}}}'
v\cdot (i{\partial}
+ {{\cal{V}}}){\widetilde{\cal{B}}}' )
\nonumber \\
& & +\left( \Delta_r-\frac{g_r f_\pi}{4}\right)
\Tr\overline{\widetilde{\cal{B}}}
 \widetilde{\cal{B}}
+\left( \Delta_r+\frac{g_r f_\pi}{4}\right)
\Tr\overline{\widetilde{\cal{B}}}'
 \widetilde{\cal{B}}'
\nonumber \\
& & -\half \Tr(\overline{\widetilde{\cal{B}}}
(v\cdot {{\cal{A}}}){\widetilde{\cal{B}}}' )
 -\half \Tr(\overline{\widetilde{\cal{B}}}'
(v\cdot {{\cal{A}}}){\widetilde{\cal{B}}} )
\nonumber \\
& & - i\frac{k_r}{2}
\Tr \overline{ \widetilde{\cal{B}}}\gamma^5 \slash{{\cal{A}}}
\widetilde{\cal{ B}}
+ i\frac{k_r}{2}
\Tr \overline{ \widetilde{\cal{B}}}'\gamma^5 \slash{{\cal{A}}}
\widetilde{\cal{ B}}'
\eea
where:
\beq
\Delta_r = \left(\Delta_{} + \frac{h_{r}}{4\Lambda}f_\pi^2\right)
\eeq
and the heavy meson
mass eigenvalues are:
\smbbo
2\Delta_r-gf_\pi/2,\;\;\makebox{and} \;\; 2\Delta_r+gf_\pi/2
\smebo
respectively (our normalization conventions
imply the physical mass shift
is $\delta M$ if
the Lagrangian contains $\half\delta M (\tr{\overline{\cal{B}}{\cal{B}}})$;
see \cite{BHHQ}). The mass eigenfields are nontrivial
functionals of the pions through the absorbed $\xi$, $\xi^\dagger$ factors.

Note the appearance of the off--diagonal pionic transition terms of the form
$\overline{\widetilde{\cal{B}}}
(v\cdot {{\cal{A}}}){\widetilde{\cal{B}}}'$.
At this stage it can be seen that these terms are associated with
a ``Goldberger--Trieman relation,'' by taking
${\cal{A}}_\mu = \partial_\mu \pi\cdot \tau/f_\pi$, integrating
by parts, and using the equations of motion.  One finds
that  the ${\cal{B}}'{\cal{B}}\pi$
amplitude has a coupling strength $ g_r$, and this is
seen to be  given by
$\Delta M/f_\pi$.\footnote{
Note that we can decouple the heavier field $\widetilde{\cal{B}}'$ to leading order
in the mass gap $g_r f_\pi$ by ``integrating it out''
which amounts to setting it to zero in leading order, equivalent to the point of departure taken by
Wise, \cite{Wise}, justified so long as $q$ is small
compared to the mass gap.}

The central observation of this analysis, however, is that the
underlying chiral representations
of the full HL meson theory is a parity doubled scheme.
The mass gap between the parity partners
arises from $VEV{\sigma}$ and leads to a 
splitting between the even and odd parity multiplets:
\smbbo
\Delta M =   g_rf_\pi
\smebo
Here $g_r$ is the $BB'\pi$
transition coupling constant and is the analogue of
the $g_{NN\pi}$ in the nucleon system.

Note that to match the formalism of III.A we have
written things in terms of $SU(2)_L\times SU(2)_R$, 
but this readily applies to $SU(3)_L\times SU(3)_R$
with the octet of Nambu-Goldstone bosons.
The axial vector current describes the main
transitions between parity partners, $(0^+,1^+)\rightarrow (0^-,1^-)+``\pi"$
where $``\pi"$ is the $SU(3)$ matching member of the octet,
e.g. $\bar{c}s\rightarrow \bar{c}u +K^-$, etc. The
 parity degeneracy of the $(0^+,1^+)$ and  $(0^-,1^-)$ states
is lifted by $\VEV{\sigma}$.

When the BABAR collaboration subsequently
reported the observation of
a narrow resonance  in $D^+_s\pi^0$ with a mass of $2317$ MeV \cite{BABAR},  
there was  also  a hint of a second state (the $1^+$) in  $D_s\pi^0\gamma$ with a mass $2460$ MeV.
The mass difference between the $D^*_s(2317)$ and the 
well established lightest charm-strange meson,
$D_s$, is $\Delta M = 349$ MeV. With this mass gap the 
decay $D_s^*(2317) \rightarrow D_{u,d}+K$ is kinematically forbidden. 

Both of these states 
are members of the heavy quark $(0^+,1^+)$ spin multiplet.  
The multiplet is then  established to a good
approximation as the {\em chiral partner} of the $(0^-,1^-)$
ground state predicted by QCD in the NJL approximation discussed above.  
The observed magnitude of $\Delta M$ 
is in remarkably good agreement with the
NJL model estimate of ref.\cite{BHHQ}.

The suppressed decays of all of these states 
must  proceeds through $SU(3)$
breaking effects, e.g., 
$D_s(2317) \rightarrow D_{u,d}+(\eta\rightarrow \pi^0)$, emitting
a virtual $\eta$ that then mixes with
the $\pi^0$ through isospin violating
effects. We have tabulated the spectrum and all decay
widths for all analogue processes in\cite{BEH}. Moreover, we
show in \cite{BEH} that electromagnetic transitions are 
indeed, and somewhat remarkably,
suppressed, since they involve cancellations between
the heavy and light magnetic moments. 

The overall picture of the chiral
structure of the heavy-light systems works
quite well.  The $D_s(2317)$ is the 
first clear observation
of a  general phenomenon that applies to the ``heavy-strange''
$(0^+,1^+)$ mesons, $\bar{b}s$ and $\bar{c}s$, 
and the $\small (\frac{1}{2}^-,\frac{3}{2}^-)$ ``heavy-heavy-strange'' baryons, 
$[cc]_{J=1}s$, $[bb]_{J=1}s$, $[cb]_{J=1}s$, and $[cb]_{J=0}s$
 \cite{BEH}.

\section{Nambu--Jona-Lasinio Inspired Composite Higgs Bosons}

\subsection{Minimal Dynamical Symmetry Breaking of Electroweak Interactions}

Prior to the top quark discovery,
following suggestions of Nambu, \cite{Nambutc}, and pioneering work of Miransky, Tanabashi and
Yamawaki, \cite{Yama},  a predictive minimal theory of a composite Higgs boson  
was constructed by Bardeen, Hill and Lindner \cite{BHL}.
This is based upon a Nambu-Jona-Lasinio model where
the Higgs is composed of $\bar{t}t$ (for a review see  \cite{CTH}\cite{NSD}). 

We assume chiral fermions, each with $N_c=3$  labeled by $(a,...)$,
where we identify $\psi_L^{a} = (t, b)_L$ and $\psi_R=t_R$
A non-confining,   $SU(3)\times SU(2)_{L}\times U(1)_{R}$ invariant
action, then takes the form:
\smbbo
\label{NJL1t}
S_{NJL}
=\!\int\! d^4x \;\bl i\bar{\psi}^{ai}_L(x)\slash{\partial}\psi_{aiL}(x)
+ i\bar{\psi}^a_R(x)\slash{\partial}\psi_{aR}(x)
\nonumbo
\qquad \qquad
+\;
\frac{g^2}{M_0^2}
\bar{\psi}^{ai}_L(x)\psi_{aR}(x)\;\bar{\psi}^b_R(x)\psi_{biL}(x)
\br.
\smebo
where $i$ is an $SU(2)$ index, summed from 1 to 2.
The interaction is factorized by a isodoublet Higgs auxiliary field
$H$, and loops are computed theory  in section \ref{Intro}, yielding
an unrenormalized action, with an induced kinetic term for $H$. 
We renormalize, $H \rightarrow \sqrt{Z}^{-1}H$, hence:

\smbbo
\label{NJL5}
\backo
S_{\mu}
=\!\int\! d^4x \;\bl 
i\bar{\psi}^a_L\slash{\partial}\psi_{aL}+ i\bar{\psi}^a_R\slash{\partial}\psi_{aR}
+
\partial_\mu H^\dagger\partial^\mu H
\nonumbo
\backo
-m_r^2H^\dagger H - \frac{\lambda_r }{2}(H^\dagger H)^2 + g_r\bar{\psi}^a_L\psi_{aR}H(x)+h.c. 
\br.
\smebo 
where the renormalized parameters are similar
to those in section \ref{Intro},
\smbbo
\label{NJL30}
\backo
m_r^2 = \frac{1}{Z}  \bl M_0^{2}\!-\!\frac{N_{c}g^{2}}{8\pi ^{2}} ( M_0^{2}-\mu^2)\br
\nonumbo
\backo
g^2_r\!=\! \frac{g^2}{Z}\!=\!\frac{8\pi^2}{N_c\ln(M_0/\mu)},\;\;
\lambda_r\!=\!\frac{\lambda}{Z^2} \!=\!\frac{16\pi ^{2}}{N_c\ln( M_0/\mu)}.
\smebo

We see that the $g_r^2$ and $\lambda_r$ each approach a Landau pole 
as the RG scale $\mu$  approaches the composite scale $M_0$.
The NJL theory thus makes two predictions:
\smbbo
(1)\qquad g^2_r\rightarrow \infty\qquad  \makebox{as} \qquad\mu\rightarrow M_0
\nonumbo 
(2) \qquad \frac{\lambda_r}{g_r^2}  \rightarrow 2 \qquad \makebox{as}\qquad \mu\rightarrow M_0
\smebo
To obtain a theory that is close to the observed top quark mass
we must tune a significant hierarchy in $M_0/m_t$,  and the logs become large
(this allows us to neglect small additive constant corrections
to the logs ).
If the logs are large, however, we must go beyond
the fermion loop approximation of NJL and include the full
renormalization group evolution, assuming the standard model.  
The remnant of the NJL dynamics is
then the compositeness conditions, eq.(50), which are boundary conditions on
the full RG.
It turns out that the solution for the renormalized Higgs-Yukawa coupling
of the minimal top condensation theory is then given by the RG ``infrared quasi-fixed point.''

Indeed, the idea that the renormalization group would yield a large top
quark mass predates the composite $\bar{t}t$ Higgs boson scheme
and was first discussed by Pendleton and Ross \cite{PR} and Hill \cite{FP}. 
The skeletal one-loop RG equation
for  the top quark Higgs-Yukawa coupling, $g_t$, is:
\beq
\label{one}
D g_t = g_t\left(\left(N_c+\frac{3}{2}\right)g_t^2 - \left(N_c^2-1\right)g_3^2\right)
\eeq
where $D= 16\pi^2 \partial/\partial \ln(\mu)$,  $\mu$ is
the running mass scale,  $g_3$ is the QCD coupling, and $N_c=3$ is the number of colors.
For illustrative purposes we suppress electroweak corrections.
Pendleton and Ross emphasized a ``tracking solution'' in which the ratio
of $g_t/g_3$ is constant.  The tracking solution
predicts a top mass of order $100$ GeV.  

In \cite{FP} the quasi-IR fixed
point was emphasized, in which the initial value of $g_t$ in the extremely far UV
is large, e.g. a Landau pole.  It was then noted that $g_t$ runs down to
a relatively constant fixed value determined by $g_3$, hence a ``quasi'' fixed point. 
This leads to a predicted top mass  $\sim 220$ GeV, as seen in Fig.(2), which depends softly upon
the initial scale $\Lambda$  as $\sim \ln(\ln(M_0/m_t)$.
The effective top mass $m_{top}=g_t(\mu) v $  (where $v = 175$ GeV is the electroweak scale)
is plotted vs. renormalization scale $\mu$ (in the figure the physical top mass corresponds $\log(\mu)\sim 2$).
Assuming the initial mass scale of the Landau pole (LP) is of order $10^{15}-10^{19}$ GeV, 
the quasi-fixed point translates 
into values of the top quark mass that are heavy, as seen in Table I. 
One obtains $m_{top} \sim 220$ GeV starting at the Planck mass.
The quartic coupling of the Higgs, $\lambda(\mu)$, is also governed 
by a similar infrared quasi-fixed point 
(see the detailed RG discussion in \cite{CTH}).
The large value of $g_t$ at $M_0$ suggests a composite Higgs boson theory.
\footnote{This is often nowadays called a ``focus point,''
but referred to as a ``fixed point'' is the traditional nomenclature inherited
from condensed matter physics.
Since a focusing occurs
in the infrared, the top quark mass
is determined to lie within a narrow range of values.}

\begin{figure}[t!]
\vspace{-1 in}
	\includegraphics[width=0.5\textwidth]{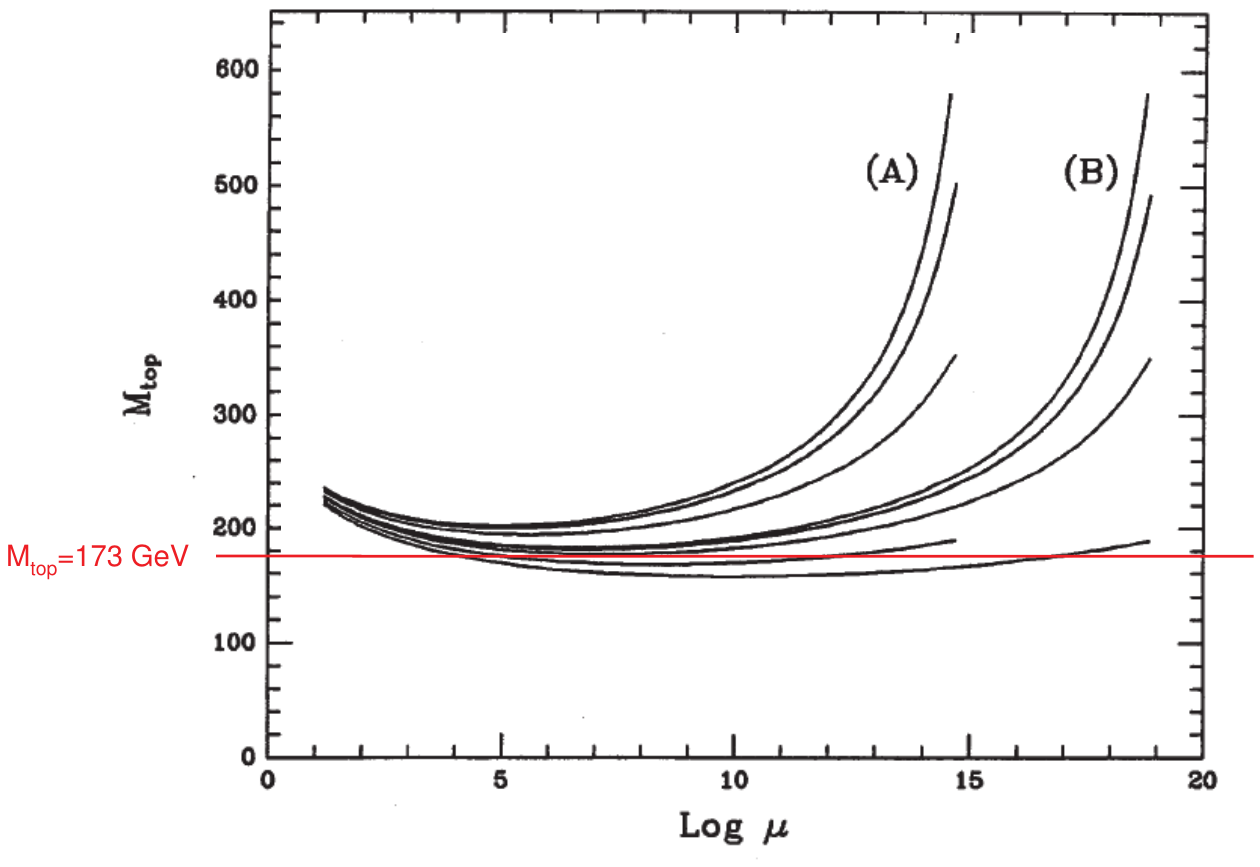}
	\vspace{-1.2 in}
	\caption{The top quark quasi--fixed point in the top mass  $m_{top}=g_tv$ where $v=175$ GeV
	is the Higgs VEV, plotted vs the running scale $\mu$, The focusing in the infrared
	and its relative insensitivity to initial values is indicated.}
	\label{RGFlowgtopSM}
\end{figure}

\begin{table}
\begin{center}
\renewcommand{\arraystretch}{2.0}
    \caption{Top quark mass for initial Landau pole $M_0$.}
    \label{table1}
    \vspace{0.05 in}
    \begin{tabular}{l|c|c|c|c|c} 
   $M_0$ GeV & $10^{19}$  & $10^{15}$  & $10^{11}$  & $10^{7}$  & $10^5$ \\
      \hline
      $m_{top}$ GeV & 220  & 230 & 250 & 290 & 360 \\
    \end{tabular}
  \end{center}
  \vspace{-0.1 in}
\end{table}

Ultimately, the top quark was discovered in 1995 at Fermilab by 
CDF and D-Zero \cite{CDF}\cite{Dzero}, with a mass of $m_t=174$ GeV.
 The top mass, though significantly larger than the original expectations,
is shy of the naive SM infrared quasi-fixed point 
with $M_0=10^{19}$ GeV by about 20\%.
I view as success of the infrared quasi-fixed point, 
given that the prediction assumed only SM
physics extending all the way up to Planck scale, however  
the whole story is yet to be unraveled.

What could bring the top mass quasi-fixed point prediction
into more precise concordance with experiment?  It turns out that it isn't easy to do this with
minor modifications of the SM.  For example,
one might consider embedding $SU(3)\rightarrow SU(3)\times SU(3)...$,
at intermediate scales in the RG running,
such as a perturbative version of 
``topcolor,'' or ``flavor universal colorons,'' or 
``latticized extra dimensions,'' etc. \cite{Topcolor}\cite{bdob1}\cite{Pokorski}. However, this
generally causes the effective $g_3$ to become larger and the quasi-fixed point 
of $g_t(m_{t})$ moves up and 
the discrepancy with experiment gets larger.

Moreover, in the $\bar{t}t$ composite  scheme the quartic Higgs coupling is too
high $\lambda \sim 1$ vs. the SM value, $\lambda\sim 0.25$,
and leads to an unacceptably large Higgs boson mass \cite{BHL}. 
This may also be informing us of required modifications of the renormalization group
beyond the simple SM inputs at higher energy scales to
bring the result into concordance with experiment.

\vspace{0.1 in}

\subsection{Scalar Democracy}

We have considered a 
maximal scalar field extension of the SM \cite{HMTT} (and a less drastic
minimal version, \cite{HMTT2}).
This was largely motivated by curiosity: How obstructive are the rare weak decay flavor constraints 
on a rich spectrum of Higgs bosons? How many scalars might exist, given the fermion
composition of the SM, and what patterns might be suggested, etc.?
Here we propose the principle that every chiral fermion pair in the SM
has a corresponding scalar bound state of the same quantum numbers.
This is suggested by thinking along the lines of the NJL model.

We call this model ``Scalar Democracy.''  We count the vertex operators
for every chiral fermion pair in the SM
and  find that
this leads to a vast spectrum of color octet isodoublets
and triplet leptoquarks, etc.  
We assume these exotic new scalar bosons have ultra-large  masses
and do not affect low energy RG running.
 The model contains eighteen Higgs doublets in the  quark sector
and likewise in the lepton sector that will have  masses
extending up to $\sim 10^5$ TeV.  These two subsectors resemble an $SU(6)_L\times SU(6)_R$
linear $\Sigma$-model Lagrangian \cite{Manohar} (as in Fig.(2)),
where the interaction is subcritical and ultimately
only the SM Higgs condenses. 

The theory has one universal HY coupling $g$ defined
at the Planck scale, which we view as the compositeness scale of all scalars.
This universal $g$ is renormalized as we flow into the infra-red  and symmetry breaking
effects lead to slightly different values in the different subsectors of the theory.
For quarks this coupling near the weak scale is identified with the top quark, $g_t=g$, 
while for leptons we obtain $g_\ell \sim 0.7 g$. 
Hence, our theory is calibrated by the known value of $g_t\approx 1$ \cite{HMTT}.
This buys us some predictivity. 

\begin{figure}[t]
\vspace{-2.2 in}
	\hspace*{-4.5cm}
	\includegraphics[angle=-90,width=1.0\textwidth]{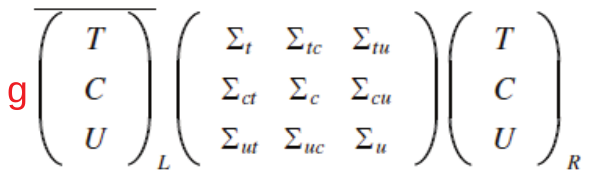}
	 \vspace{-2.0 in}
	 \vspace{-0.5 cm}
	 \caption{ 
	 $SU(6)_L\times SU(6)_R$ chiral Lagrangian structure of the scalar democracy 
	 in the quark sector.  $T=(t,b)$, $C=(c,s)$, $U=(u,d)$ and $\Sigma_{xy}=(H_x, H_y^C)$.
	The quark Higgs subsector becomes a subcritical $SU(6)_L\times SU(6)_R$
linear $\Sigma$-model Lagrangian, where $\Sigma$ is a $6\times 6 $ complex matrix that can be viewed
as nine $2\times 2$ complex $\Sigma$ fields where each $\Sigma_{xy}=(H_x, H^C_y)$
is a pair of Higgs doublets. There is one universal coupling $g$ which is subject to
RG effects.}
	\label{fig:Ft_regions}
\end{figure}

Here
 the  flavor physics and mass hierarchy problems are flipped 
 out of the Higgs-Yukawa coupling texture  and into
the mass matrix of the many Higgs fields. 
The Higgs mass matrix is input as $d=2$ gauge invariant operators. We have no theory
of these, but we must choose the inputs to fit the quark and lepton sector masses
and CKM physics, as  well as maintain consistency with rare weak decays, etc.  
It is not obvious {\em a priori} that there exists a consistent solution with the flavor 
constraints.

The RG equation for the universal
HY coupling  in the quark subsector now takes the one-loop form
\cite{HMTT}:
\beq
\label{two}
D g_t = g_t\left((N_c+N_f)g_t^2 - (N_c^2 -1)g_3^2\right)
\eeq
Note the enhanced coefficient of $g_t^2$ in eq.(\ref{two})
relative to eq.(1) where $N_f$ is the number
of flavors, i.e., $N_f=6$ in the quark sector of the model.

\begin{figure}[t!]
\vspace{-1 in}
	\includegraphics[width=0.5\textwidth]{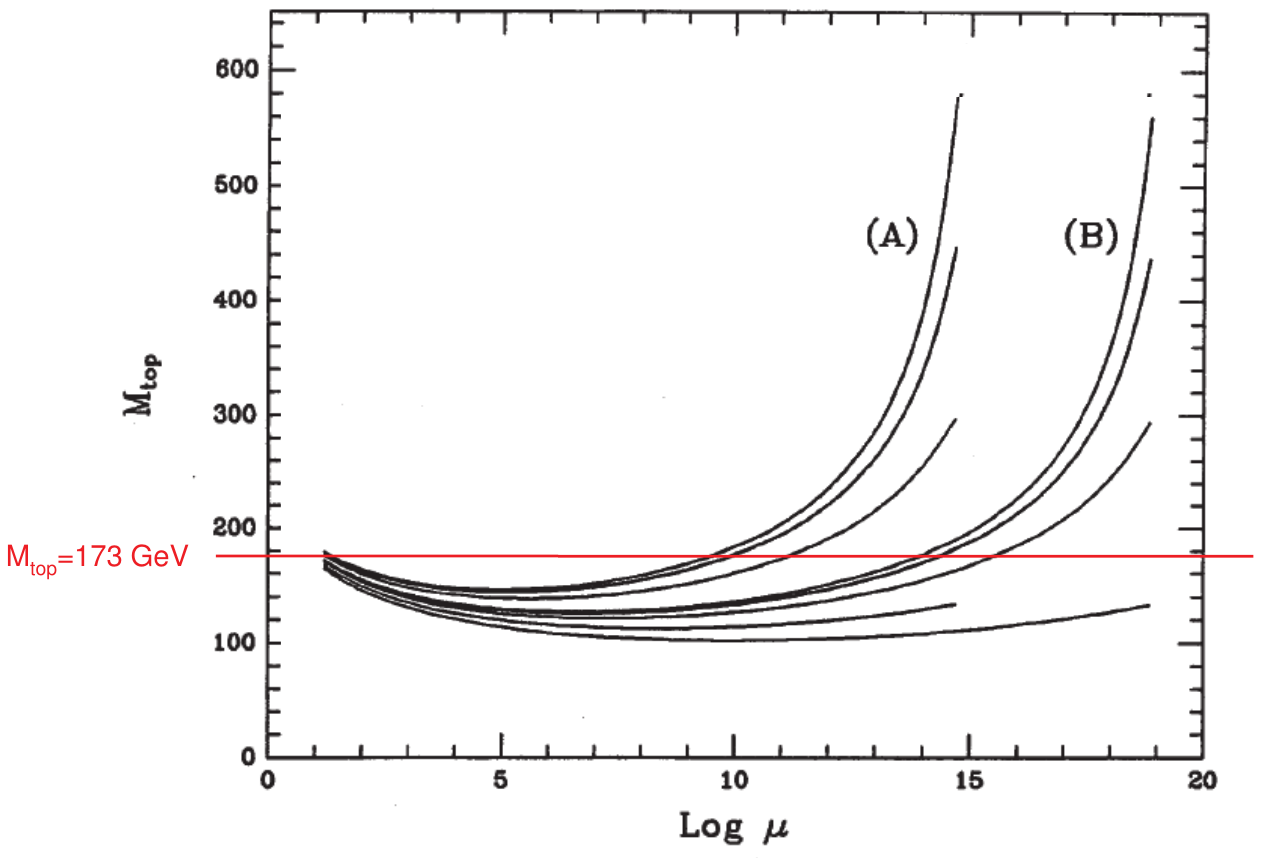}
	\vspace{-1.2 in}
	\caption{The top quark fixed point is shifted down in the $SU(6)\times SU(6)$
	model and becomes concordant with the observed top mass, with residual
	corrections coming from sensitivity to the extended Higgs spectrum (see Table II). }
\end{figure}

\begin{table}
\begin{center}
\renewcommand{\arraystretch}{2.0}
    \caption{Top quark mass for initial Landau pole $M_0$.}
    \label{table2}
    \vspace{0.05 in}
    \begin{tabular}{l|c|c|c|c|c} 
   $M_0$ GeV & $10^{19}$  & $10^{15}$  & $10^{11}$  & $10^{7}$  & $10^5$ \\
      \hline
      $m_{top}$ GeV & $156+\Delta$  & $162+\Delta$ & $177+\Delta$ & $205+\Delta$ & $254+\Delta$ \\ 
    \end{tabular}
  \end{center}
  \vspace{-0.1 in}
\end{table}

The resulting RG evolution is shown in Fig.(4) 
and the top quark mass predictions are given in Table II.
There is residual sensitivity to the decoupling scale of the heavy Higgs bosons, 
summarized by the parameter $\Delta$, which 
reduces the effective $N_f$ below the Higgs decoupling scale as discussed in
\cite{HMTT}\cite{HMTT2}. There we estimate  $\Delta \approx
(2.8\;GeV)\ln(\langle M_H \rangle/10^2)$ hence $\Delta \approx 19$ GeV for an average
heavy doublet mass $ M_H \approx 100$ TeV.  Hence, the top mass can be accommodated
for $M_X \sim 10^{19}$ GeV, and is indirectly probing the new physics of the additional
heavy Higgs bosons.

\begin{table*}
\label{diagonal}
\centering
\renewcommand{\arraystretch}{1.3}
\begin{tabular}{|p{3cm}p{4.1cm}p{3cm}l|} 
	\hline
	Higgs field & Fermion mass &  Case (1) [TeV]  &  Case (2) [TeV] \\
	\hline
	\hline
        $H^{\prime}_0 = v + \frac{h}{\sqrt{2}}$ & $m_t = gv=175$ GeV  & $m_H=0.125$  &  $m_H=0.125$ \\
	\hline
	$H^{\prime}_{b}=v\frac{\mu^2}{M_b^2} + H_{b}$ & $ m_b= gv\frac{\mu^2}{M_b^2}=4.5$ GeV   & $M_b=3.5$  &  $M_b=0.620  $     \\
	\hline
	$H^{\prime}_{\tau}=v\frac{\mu^2}{M_\tau^2} + H_{\tau}$ &  $ m_\tau= g_\ell v\frac{\mu^2}{M_\tau^2}=1.8$
	GeV &  $M_\tau= 6.8 $   &  $M_\tau= 0.825 $  \\
	\hline
	$H^{\prime}_{c}=v\frac{\mu^2}{M_b^2} + H_{c}$ &  $ m_c= gv\frac{\mu^2}{M_c^2}=1.3$ GeV &  $M_c=13.5$  &  $M_c=1.2$  \\
	\hline
	$H^{\prime}_{\mu}= v\frac{\mu^2}{M_\mu^2} + H_{\mu}$ &  $ m_\mu= g_\ell v\frac{\mu^2}{M_\mu^2}
	=106$ MeV &  $M_\mu= 1.2\times 10^2$  & $M_\mu= 3.4$   \\
	\hline
	$H^{\prime}_{s}=v\frac{\mu^2}{M_s^2} + H_{s} $ & $m_s= gv\frac{\mu^2}{M_s^2}=95$ MeV & $M_s= 1.8\times 10^2$  & $M_s= 4.3 $  \\
	\hline
	$H^{\prime}_{d}=v\frac{\mu^2}{M_d^2} + H_{d} $ &  $m_d= gv\frac{\mu^2}{M_d^2}=4.8$ MeV 
	& $M_d=3.6\times 10^3  $ & $M_d=19$  \\
	\hline
	$H^{\prime}_{u}=v\frac{\mu^2}{M_u^2} + H_{u} $ &  $m_u= gv\frac{\mu^2}{M_u^2}=2.3$ MeV 
	& $M_u= 7.6\times 10^3 $  & $M_u= 27 $ \\
	\hline
	$H^{\prime}_{e}= v\frac{\mu^2}{M_e^2} + H_{e}$ &   $m_e= g_\ell v\frac{\mu^2}{M_e^2}=0.5$ MeV
	& $M_e=2.45\times 10^4$  &  $M_e=49$ \\
	\hline
\end{tabular}
\caption{The estimates for diagonal Higgs bosons masses 
in the limit of no CKM mixing, and assuming 
(1) the level-repulsion feedback on the 
Higgs mass term is limited to $(100 \,\text{GeV})^2$  for each of the quarks and leptons, hence $M_q= (100$ GeV$)(m_t/m_q )$ and 
$M_\ell= (100$ GeV$)(g_{\ell}m_t/m_{\ell})$.
(2) $\mu=100$ GeV for all mixings, hence $M_q= \mu (m_{t}/ m_{q})^{1/2}$  and $M_\ell= \mu (m_{t}g_{\ell}/gm_{q})^{1/2}$.
Here  $g=1$, $g_\ell=0.7$ and $ v= 175 $ GeV. For a more detailed discussion see \cite{HMTT}. 
\label{tab:Higgs_masses}
}
\label{tab:Higgs_mass_est}
\end{table*}

 The CKM physics is generated by off-diagonal Higgs field and their mixings.
We ran numerous benchmark tests of hypothetical Higgs mass matrices to
study consistency with flavor physics (rare weak decay constraints)
and the generation of quark and lepton masses and some of these are discussed and displayed  in  
is \cite{HMTT,HMTT2}.
The results are nontrivial, yet
we find there are consistent solutions where the
masses and mixings of a spectrum of new Higgs doublets can explain the
entire fermionic mass and CKM-mixing structure of the SM
with the single universal HY coupling.
Some results for mass limits are quoted in  Table III.

The generic idea of a maximally large spectrum of Higgs bosons, 
``scalar democracy,'' is proposed in \cite{HMTT,HMTT2}. 
Related prior literature includes 
models of Ben Tov, Porto, and Zee, et.al., \cite{Zee}\cite{Bentov} which considers only
diagonal Higgs fields (typically $6$ in each subsector) and 
expects new Higgs states in the TeV mass range. 
Related enhanced spectroscopy of composite Higgs bosons arises in various
other models, e.g.,
\cite{Marco}\cite{Osipov}\cite{Rode}\cite{Ishida}.
A similar large number of Higgs doublets
has been invoked in an interesting  model of Weinberg's asymptotic safety \cite{San}.
This may preclude a conventional GUT picture if the unification of gauge interactions
occurs in tandem with the non-perturbative asymptotically safe fixed point.
Scalars go hand-in-hand with Weyl invariance, and recent  models
suggest a unified picture of inflation occurring with the dynamical generation of
the Planck scale \cite{Weyl}, which may provide a larger context for this scenario.
Also, the idea of universal large Higgs-Yukawa coupling, e.g.,  $H_b$ with $g=1$, 
was previously introduced in the context of a Coleman-Weinberg Higgs potential \cite{CWH}
(see also \cite{Rao}\cite{thesis}).

\subsection{Counting Scalars \label{cs}}

If we assume one bound state in nature per $s$-wave fermion pair at some high
scale $M_X$, then we can count the number of composite scalars.
All of the SM matter fields can be represented by $48$ two-component left-handed 
spinors, $\psi _{A}^{i}$. 
This includes all the left-handed and anti-right-handed fermions.
We can collect these into a large global $SU(48)\times U(1) $ multiplet, 
the new dynamics that is blind to the SM gauge interactions.
We emphasize that this is a dynamical symmetry, and familiar GUT 
theories that contain only the SM fermions will be gauged subgroups of this $SU(48)$.
Here the indices $\left( i,j\right) $ run over all the $48$ flavor, doublet, 
and color degrees of freedom of the SM fermions.

The most general non-derivative ($s$-wave) scalar-field  bilinear 
we can construct of these fields is through the ``vertex operator,'' $\epsilon^{AB}\psi_{A}^{i}\psi_{B}^{j}$:
\bea
\epsilon^{AB}\psi_{A}^{i}\psi_{B}^{j}\Theta_{ij}+h.c.,
\eea
where $\Theta _{ij}$ transforms as the symmetric $ \mathbf{1176} $ representation of  $SU(48)$ 
(this is analogue to the sextet representation of $ SU(3) $). 
The field $\Theta _{ij}$ contains many complex scalar fields with assorted quantum numbers, 
including baryon and lepton number, color, and weak charges.

To make contact with the SM fields, we consider the usual $24$ 
left-handed quarks and leptons, $\Psi_{Li}$, and the $24$ right-handed counterparts, $\Psi_{R\widehat{i}} $.
 The index $i$ now runs over the chiral $SU(24)_{L}$ and $\widehat{i}$ over the chiral $SU(24)_{R} $ subgroups of $ SU(48) $. We thus have:
\bea
\label{fields}
\Phi_{i\widehat{j}}\overline{\Psi }_{L}^{i}\Psi _{R}^{\widehat{j}}+\Omega
_{ij}\overline{\Psi }_{L}^{i}\Psi _{R}^{jC}+\widehat{\Omega }_{\widehat{ij}%
}\overline{\Psi }_{R}^{\widehat{i}}\Psi _{L}^{\widehat{j}C}+\text{h.c}.,
\eea
where $ \Phi_{i\widehat{j}} $ is the $ (\mathbf{24}_L,\, \mathbf{24}_R) $ complex scalar field with $24^{2}=576$ complex degrees of freedom. 
$\Omega $ and $ \widehat{\Omega } $ are the symmetric $\mathbf{300} $ representation of $SU(24)_{L}$ and $SU(24)_{R}$ respectively
 These match  the degrees of freedom of $\Theta _{ij}$.
Here $\Omega _{ij}$ and $\widehat{\Omega }_{ij}$ are the analogues of Majorana masses and carry fermion number, while $\Phi $ contains fermion number neutral fields, such as Higgs fields, in addition to ($B-L$) leptoquark multiplets.

The $\Phi ,\Omega$ and $\widehat{\Omega }$ fields can be viewed as the ``composite fields'' 
arising from a NJL model effective description of the new forces.
Consider just the $SU(24)_{L} \times SU(24)_{R}\times U(1) \times U(1)_{A}$ invariant NJL model:
\bea
\label{eq:NJL}
- \frac{g^{2}}{M^{2}} (\overline{\Psi}_{L}^{i} \Psi _{R}^{j}) (\overline{\Psi}_{R,i} \Psi_{L,j}),
\eea
where the negative sign denotes an attractive interaction in the potential.
It should be noted that we can equally well write current-current (and tensor-tensor) 
interactions, mediated by 
heavy spin-1 bosons (or Pauli-Fierz spin-2 gravitons); 
these will generally contain scalar
channels and will Fierz rearrange to effectively reduce to eq.(\ref{eq:NJL}) with the attractive signs.
There also exists the possibility of the following NJL models:
\bea
-\frac{g^{2}}{M^{2}} (\overline{%
\Psi }_{L}^{i} \Psi_{R}^{Cj} ) (\overline{\Psi}_{R,i}^{C} \Psi_{L,j} ) \;\;
\makebox{or} \;\; \left( R \leftrightarrow L\right),
\eea
which lead to the composite bosons $\Omega $ and $\widehat{\Omega}$. 
The first step to solving an NJL theory would be to factorize the interaction of eq.(\ref{eq:NJL}) by introducing auxiliary scalar fields.
This leads to the equation we started with,  eq.(\ref{fields}), where  $\Phi ,\Omega$ and $\widehat{\Omega} $ are auxiliary fields.

A universal flavor and color blind interaction will bind fermion pairs into scalars 
that are bound states of ordinary quarks and leptons and will generate a plethora of Higgs doublets. 
E.g., for any fermion pair we expect there is a mini-black-hole 
with mass of order $M_{Planck}$ \cite{BHole}.
These bound states will have a universal Yukawa coupling $g$ at the scale $M^{2}$.
With $g$ taking on a near-but-subcritical value, these bound states would
have large positive masses but can be tuned to be lighter than $M$.
Symmetry-breaking effects are required to split the spectroscopy, including the 
SMH down to its observed negative mass term.
All other doublets remain heavy, but will mix. 
There is fine-tuning required to engineer the light $M^2$ as $d=2$ operators, however many
of these terms are technically natural, protected by the $SU(48)$ symmetry structure
(see \cite{HMTT} for further details).

\subsection{The Top-Bottom Subsector}

What might convince us that an extended Higgs boson spectrum exists in nature?  
The discovery of the first sequential Higgs boson, the $H_b$ with
a coupling $g\approx g_t\sim O(1)$ to $\bar{b}b$ would 
lend some credibility to the scenario.
 $H_b$  has an upper bound
on its mass of about $5.5$ TeV based on a ``no unnatural cancellations'' hypothesis. It may be
discovered up to 3 TeV with the LHC, and conclusively so with the energy doubled LHC.  
Indeed, the LHC already has the capability of placing useful
limits on the $H_b$.

The top-bottom subsystem is a subsector of the full scalar democracy
which can be defined in a self-contained way.  It can also be
easily extended to include the $(\tau, \nu_\tau)$ leptons, and we will only quote results \cite{HMTT2}.
We presently assume the top-bottom subsystem is approximately
invariant under a simple extension of the Standard Model symmetry group
structure
\beq
\label{G}
G = SU(2)_L\times SU(2)_R\times U(1)_{B-L}\times U(1)_A.
\eeq
To implement $G$ in
the $(t,b)$ sector we require that the SM Higgs doublet,
$H_0$, couples to $t_R$ with coupling  $y_t\equiv g_t$ in the usual way, and a second Higgs doublet, $H_b$,
couples to $b_R$ with coupling $g_b$. The symmetry $G$ then dictates 
that there is only a single 
Higgs-Yukawa coupling $g= g_t =g_b$ 
in the quark sector.  This coupling is thus determined by
the known top quark Higgs-Yukawa coupling, $y_t =g \simeq 1$.

The assumption of the symmetry, $G$, of Eq.~(\ref{G}) for the top-bottom system
leads to a Higgs-Yukawa (HY)
structure that is reminiscent of  the chiral Lagrangian of the proton and neutron
(or the chiral constituent model of up and down quarks \cite{Manohar};
in a composite Higgs scenario  this model was considered
by Luty \cite{Luty}):
\bea\label{one}
V_{HY} = g\overline{\Psi}_L\Sigma\Psi_R + h.c.\quad
\text{where}
\quad
\Psi= \left( \begin{array}{c}
t \\ 
b%
\end{array}%
\right).
\eea 
$\Sigma$ can be written
in terms of two column doublets,
\beq
\label{sigma}
\Sigma=(H_0,H^c_b),
\eeq
where  $H^c=i\sigma_2 H^*$. Under $SU(2)_L$ we have 
$H\rightarrow U_L H$s.
The HY couplings of Eq.~(\ref{one}) become
\bea
V_{HY} = g_t \left( 
\bar{t}, \, \bar{b}
\right)_{L} H_{0} t_{R} +g_b \left( 
\bar{t}, \, 
\bar{b}%
\right)_{L} H^c_{b} b_{R} + h.c.,
\eea
where the $SU(2)_R$ symmetry has forced $g= g_t= g_b$.

Given Eqs.~(\ref{one})and (\ref{sigma}), 
we postulate a new potential,
\bea
\label{potential1}
V&=&
M_1^2 Tr\left( \Sigma ^{\dagger}\Sigma \right)
-M_2^2 Tr\left( \Sigma ^{\dagger }\Sigma \sigma_3 \right)
\\ 
&& \!\!\!\!\!\!\!\!\!    \!\!\!\!\!  +\mu ^{2}\left(e^{i\theta} \det\Sigma + h.c.\right)
+ \frac{\lambda_1 }{2} Tr\left( \Sigma ^{\dagger }\Sigma\right)^{2}
+ \lambda_2 |\det\Sigma|^2.
\nonumber 
\eea
Using Eq.~(\ref{sigma}), $V$ can be written in terms of $H_0$ and $H_b$:
\bea
\label{potb}
V &=& 
M_{H}^{2}H_{0}^{\dagger }H_{0}+M_{b}^{2}H_{b}^{\dagger }H_{b}
+\mu^{2}\left( e^{i\theta} H_{0}^{\dagger }H_{b} + h.c. \right) 
\nonumber \\
&& +\frac{\lambda }{2}\left(
H_{0}^{\dagger }H_{0}+H_{b}^{\dagger }H_{b}\right) ^{2}+\lambda ^{\prime
}\left( H_{0}^{\dagger }H_{b}H_{b}^{\dagger }H_{0}\right),
\eea
where $\lambda_2=\lambda' + \lambda$, $M_H^2=M_1^2-M_2^2$, and  $M_b^2=M_1^2+M_2^2$.
In the limit $M_b^2\rightarrow \infty$ the field $H_b$  decouples,
and Eq.~(\ref{potb}) reduces to the SM  if
$M_H^2 \rightarrow M_0^2$, and $\lambda \rightarrow 0.25$.

We assume the quartic couplings are all of order the SM  value $\lambda\sim 0.25$.
{ Those associated with the new heavy Higgs bosons, such as $\lambda'\sim \lambda$, 
will therefore contribute negligibly small effects since they involve 
large, positive, $M^2$ heavy Higgs fields. We also set $\theta=0$.} 

Then, varying the potential
with respect to $H_b$, 
the low momentum components of $H_b$ become locked to $H_0$:
\beq
H_{b}= -\frac{\mu^2}{M_b^2}  H_{0}+ O(\lambda, \lambda').
\eeq
Substituting back into $V$ we recover the SMH potential
\bea
V &=& 
 M_0^2 H_{0}^{\dagger }H_{0}
 +\frac{\lambda }{2}\left(
H_{0}^{\dagger }H_{0}\right)^{2}+O\left(\frac{\mu^2}{M_b^2}\right),
\eea
with
\beq
\label{M0}
M_0^2= M_{H}^{2}-\frac{\mu^4}{M_b^2}.
\eeq
Note that, even with $M_H^2$ positive, $M_0^2=-(88.4)^2$ GeV$^2$
can be driven to its negative value by
the mixing with $H_b$ (level repulsion).
We minimize the SMH potential and define 
\bea
\label{SMHpot}
H_{0}=\left( 
\begin{array}{c}
v +\frac{1}{\sqrt{2}}h\\ 
0
\end{array}%
\right), \qquad v=  {174}\;{GeV},
\eea
in the unitary gauge. 
The minimum of Eq.~(\ref{SMHpot}) 
yields the usual SM result
\beq
v^2 = -M_0^2/\lambda, \qquad m_h  = \sqrt{2}|M_0|=125\;\makebox{GeV},
\eeq
where $m_h$ is the propagating Higgs boson mass.
We can then write
\beq
H_b\rightarrow H_b -\frac{\mu^2}{M_b^2}\left( 
\begin{array}{c}
v +\frac{1}{\sqrt{2}}h\\ 
0
\end{array}%
\right),
\eeq
for the full $H_b$ field. 
This is a  linearized (small angle) approximation to the mixing,
and is reasonably insensitive to the small $\lambda, \lambda'\ll1$.

We have exploited the effect of ``level repulsion'' of the Higgs mass, $M_0^2$,
downward due to the mixing with heavier $H_b$.
The level repulsion in  the presence of $\mu^2$ and $M_b^2$
occurs due to an approximate ``seesaw'' Higgs mass matrix
\beq
\label{matrix}
\begin{pmatrix}
	M_H^2 & \mu^2 \\
	\mu^2 & M_b^2
\end{pmatrix}.
\eeq
The input value of the mass term $M_H^2$ is
unknown and in principle arbitrary, and can 
have either sign. 
Let
us consider the case  $M_b^2>>M_H^2$.
Then  Eq.~(\ref{matrix}) has eigenvalues
$M_0^2=-\mu^4/M_b^2$, and $ M_b^2$.
Thus, in the limit of small, nonzero $|M_H^2|$,
we see that a negative $M_0^2$ arises naturally, and to a good approximation the physical Higgs mass 
is generated entirely by this negative mixing term.

The $b$-quark then receives its mass from $H_{b}$,
\beq
\label{mub1}
m_b =  g_b(m_b)v\frac{\mu^{2}}{M_b^{2}}
=m_{t}\frac{g_b(m_b)\mu^{2}}{g_t(m_t)M_b^{2}},
\eeq
where we have indicated the renormalization group (RG) scales 
at which these couplings should be evaluated.

In a larger scalar democracy framework 
both $g_t(m)$ and $g_b(m)$ have 
a common renormalization group equation modulo $U(1)_Y$ effects.
This implies 
$g(M_P)= g_t(M_P)=g_b(M_P) \gta 1$. Then,
we will predict 
$g(M_b)\simeq  g_t(M_b)\simeq g_b(M_b)$,  e.g., where all the values 
at the mass scale $M_b$ are determined by the RG fixed point.  

Furthermore, we find that $g_t(m)$, and more so $g_b(m)$, increase somewhat as we evolve 
downward from $M_b$ to $m_t$ or $m_b$. The top quark mass is then $m_t=g_t(m_t) v$ 
where $v$ is the SM Higgs VEV.  From these effects we obtain
the ratio
\beq
\label{Rtau}
R_b = \frac{g_b(m_b)}{g_t(m_t)} \simeq 1.5.
\eeq
The $b$-quark then receives its mass from the tadpole VEV of $H_{b}$.
\beq 
\label{mub}
m_b = g_b(m_b) v\frac{\mu^{2}}{M_b^{2}}=m_{t} R_b\frac{\mu^{2}}{M_b^{2}}.
\eeq
In the case that the Higgs mass, $M_0^2$,
is due entirely to the level repulsion by $H_b$,  i.e.. $M_H^2 =0$, 
(see also \cite{Ishida} and references therein)
using Eqs.~(\ref{M0}), (\ref{Rtau}) and (\ref{mub}), we obtain
a predicted mass of the $H_b$, 
\beq
M_b = \frac{m_{t}}{m_b} R_b|M_0|  \simeq   {5.5}\;{TeV},
\eeq
with $m_b =  {4.18}\;{GeV} $, $ m_t= {173}\;{GeV}$, and $|M_0| =  {88.4}\;{GeV}$.
We have
 ignored the effects of the quartic couplings $\lambda$,
which we expect are small.

This is a key prediction of the model.  In fact, we can argue
that with $M_H^2$ nonzero, but with small fine-tuning the result
$M_b\lta  {5.5}\;{TeV}$ is obtained.
This mass scale is accessible to the LHC
with luminosity and energy upgrades, and we feel represents an important
target for discovery of the first sequential Higgs Boson.

This simple $(t,b)$ system described above can be extended to the third generation
leptons $(\nu_\tau, \tau)$. 
Remarkably the predictions for the mass spectrum
are sensitive to the mechanism of neutrino mass generation.
The next sequential massive Higgs iso-doublet, in addition
to $H_b$, is likely to include an $H_\tau$,
and possibly also $H_\nu$, which is dependent
upon whether neutrino masses are Majorana
or Dirac in nature.   If neutrino masses are Dirac
then $H_\nu$ is very heavy, $\sim 10^{16}$ GeV, and then
have a Dirac seesaw and we can ignore $H_\nu$.

The details of this and phenomenology of the sequential Higgs bosons are discussed in 
 \cite{HMTT,HMTT2}. 
Our estimated discovery luminosities
for $h^0_b $ are seen to be attainable at the LHC or its upgrades. Moreover, the lower mass range $\lesssim 1$ TeV
is currently within range of the LHC and collaborations should attempt to place
limits.

\section{Construction of Bilocal Composites in a Local Scalar Field Theory  $\label{bloc}$ } 

The NJL model is a point-like interaction generating a point-like bound state
in field theory.  However, realistic theories, such as QCD or electrodynamics, feature
extended interactions and non-point-like bound states.  
What happens if we extend the NJL model to a non-point-like UV completion
theory?  How can we formulate the formation of non-point-like
bound states, yet remain close to NJL to garner intuition?
This section is based upon ref.\cite{HComp}.

Many years ago Yukawa proposed a multilocal field theory
for the description of relativistic bound states 
\cite{Yukawa}.  
For a composite scalar field, consisting of a pair of constituents,
he introduced a complex bilocal field, $\Phi(x,y)$. This is
factorized,  $\Phi(x,y)\rightarrow \chi(X)\phi(r)$
where $X^\mu=(x^\mu\!+\!y^\mu)/2$ where $r^\mu=(x^\mu\!-\!y^\mu)/2$, and $\chi(X)$ 
 describes the center-of-mass 
motion like any conventional point-like field. Then, $\phi(r)$ 
describes the internal structure of the bound state.
The formalism preserves Lorentz
covariance, though we typically ``gauge fix'' to the center-of mass frame
and Lorentz covariance is not then manifest. Here we must
confront ths issue of ``relative time.'' 

Each of the
constituent particles in a relativistic bound state carries its own 
local clock, e.g., $x^0$ and $y^0$. These are in principle independent, so the question
arises, ``how
can a description of a multi-particle bound state be given 
in a quantum theory with a single time variable, $X^0$?''
Yukawa introduced an imaginary ``relative time'' $r^0=(x^0-y^0)/2$, 
but this was not effective and abandon it.

 We show that a bilocal field theory 
formalism can be constructed in an action  from
a local constituent field theory. 
The removal
of relative time is associated
with the canonical normalization
of the constituent fields $\chi$ and $\phi$.
In the center-of-mass frame
the internal wave-function reduces to a static field, $\phi(\vvr)$, where $\vvr=(\vvx-\vvy)/2$.
This leads to a fairly simple
solution to the problem of  relative time,
matching the conclusions one gets from the 
Dirac Hamiltonian constraint theory \cite{Dirac}\cite{Reltime}.
For us, the resulting $\phi(\vvr)$ appears as a  straightforward  result.

After first considering a bosonic construction,  we apply this to
a theory of chiral fermions with an extended interaction mediated by
a perturbative massive gluon, i.e., the ``coloron model'' 
\cite{Topcolor}\cite{Bijnens}\cite{Simmons}\cite{NSD}.
This provides a UV completion for 
the Nambu--Jona-Lasinio  model, recovered in the point-like
limit, $\vvr \rightarrow 0$ and
leads to an effective (mass)$^2$ Yukawa potential with coupling $g$.
We  form bound states with mass, 
$m^2$, determined as the eigenvalue of a static Schr\"odinger-Klein-Gordon (SKG) equation for
the internal wave-function $\phi(\vvr)$. 

A key result of this analysis is that
the coloron model has
a nontrivial {\em classical critical behavior}, $g>g_c$, 
leading to a bound state with a negative $m^2$.  The classical interaction
is analogous to the Fr\"ohlich Hamiltonian interaction in a superconductor and
has a BCS-like enhancement of the coupling by a factor of $N_c$ (number of colors) \cite{Cooper}.
Remarkably we find the classical $g_c$ is numerically close to the NJL critical coupling constant
which arises in fermion loops. 
Moreover, the scalar bound state will have an effective Yukawa coupling to its constituent fermions,
distinct from $g$, 
that is emergent in the theory. 

The description of a relativistic bound state in the rest frame is 
similar to the eigenvalue problem of the nonrelativistic Schr\"odinger equation
and some intuition carries over.  However,
the eigenvalue of the static Schr\"odinger-Klein-Gordon (SKG) 
equation, is $m^2$, rather than energy. Hence, a 
bound state with positive $m^2$  is a resonance that can decay to its
constituents and has a Lorentz line-shape in $m^2$ (we give an example in
\cite{HComp}), and thus has a large
distance radiative component in its solution
that represents incoming and outgoing open scattering states. 

If the eigenvalue for $m^2$ is negative, or tachyonic, then we have 
chiral vacuum instability, which requires a
quartic interaction of the composite field, $\sim \lambda \Phi^4$.  
In the broken symmetry phase the composite field $\Phi(x,y)$ acquires a vacuum expectation value (VEV),
$\langle\Phi\rangle=v$. 
In the perturbative ($\lambda$) solution, 
in the broken phase $\phi(\vvr)$ remains localized 
and the Nambu-Goldstone modes and Brout-Englert-Higgs (BEH)
boson retain the common localized solution for their internal wave-functions.

\subsection{Scalar Theory}

Consider local scalar fields $\varphi(x)$ (complex) and $A(x)$ (real) and action:
 \smbbo 
 \label{start}
\backo
S=\int_x\bl |\partial\varphi|^2 \!+\! \half (\partial A)^2\!-\!\half M^2A^2\!-\!gM|\varphi|^2A
\!-\!\frac{\lambda}{2}|\varphi|^4\br,
\nonumbo
 \smebo
 where we abbreviate, $|\partial\varphi|^2=\partial_\mu\varphi^\dagger\partial^\mu\varphi$
 and $(\partial A)^2= \partial_\mu A\partial^\mu A$.
Here $g$ is dimensionless and we refer all mass scales to the single scale $M$. 
We will discuss the quartic term separately below, and presently set it aside, $\lambda=0$.

If we integrate out $A$ we obtain an effective, attractive, bilocal potential interaction term
at leading order in $g^2$, 
 \smbbo
\label{act00}
\backo\backo
S=\!\!\int_x\! |\partial\varphi|^2 
+\frac{{g^2M^2}}{2}\!\!\!\int_{xy}\!\!\!\! \varphi^\dagger\!(y)\varphi(y)D_F(y\!-\!x)
\varphi^\dagger\!(x)\varphi(x),
 \smebo
where the two-point function is given by $(i)\times$ the Feynman propagator, 
 \smbbo
 \label{DF}
D_F(x-y)=-\!\!\int \frac{e^{iq_\mu(x^\mu-y^\mu)}}{(q^2-M^2)} \frac{d^4q}{(2\pi)^4}.
 \smebo
In the action eq.(\ref{act00}), the kinetic term is still
local while the interaction is bilocal, and the theory is still classical in
that this only involved a tree diagram that is  ${\cal{O}}(\hbar^{0})$.

We now define a bilocal field of mass dimension $d=1$,
 \smbbo 
\label{biloc}
\Phi(x,y)=M^{-1}\varphi(x)\varphi(y) .
 \smebo
The free particle states described by the bilocal field trivially satisfy
an equation of motion,
\smbbo
 \label{eq1}
 \partial_x^2\Phi(x,y)=0 ,
 \smebo
 and this is generated by a bilocal action,
 \smbbo
\label{act0}
\backo
S=M^4\!\!\int_{xy}\!\! Z|\partial_x\Phi(x,y)|^2,
\smebo
where we will specify the normalization, $Z$ and scale $M$ subsequently.
With the bilocal field the interaction of eq.(\ref{act00}) becomes,
 \smbbo
 \frac{{g^2M^4}}{2}\!\!\!\int_{xy}\!\!\!\! \Phi^\dagger(x,y)D_F(x\!-\!y) \Phi(x,y).
 \smebo

We can therefore postulate a bilocalized action as a free particle
part plus the interaction,
\smbbo
\label{act0}
\backo
S=\!\int_{xy}\!\!\bl ZM^4|\partial_x\Phi(x,y)|^2 \!
\nonumbo \qquad
+\!\half{g^2M^4}\Phi^\dagger(x,y)D_F(x-y)\Phi(x,y)\br.
\smebo
In the limit $g=0$ the field $\Phi(x,y)$ and the action faithfully represents
two particle kinematics.

We see that a $U(1)$ conserved Noether current is
generated by $\Phi(x,y)\rightarrow e^{i\theta(x)}\Phi(x,y)$,
\smbbo
J_{\Phi\mu}(x)= 
iZ\int d^4y\bl \Phi^\dagger (x,y) \frac{\stackrel{\leftrightarrow}{\partial}}{\partial x^\mu}\Phi(x,y)\br
\smebo
where $A\stackrel{\leftrightarrow}{\partial}B=
A{\partial}B-({\partial}A)B$. We can demand this match the conserved $U(1)$  current in
the constituent theory, 
\smbbo
J_{\varphi \mu}(x)= 
i\varphi^\dagger (x) \frac{\stackrel{\leftrightarrow}{\partial}}{\partial x^\mu}\varphi(x)
\smebo
Substituting  eq.(\ref{biloc}) into $J_{\Phi\mu}(x)$ we see that the matching requires
 \smbbo
\label{match}
J_{\Phi\mu}(x)=J_{\varphi \mu}(x)\; ZM^{2} \!\!\int \!d^4y |\varphi(y)|^2
 \smebo
hence
 \smbbo
\label{norm000}
1=
ZM^{2}\int \!d^4y \;|\varphi(y)|^2
 \smebo
 This is a required constraint for the bound state sector of the theory.
  Note that the square of the constraint is the 4-normalization of $\Phi$ 
  \smbbo
\label{normphi}
\backo
1=
Z^2M^{4}\int \!d^4y\; d^4y \;|\varphi(y)|^2|\varphi(x)|^2
\nonumbo
=
Z^2M^6\int \!d^4y\;d^4y \;|\Phi(x,y)|^2
 \smebo
 This implies that the presence of a correlation in the two particle sector, $\Phi(x,y)$
 acts as a constraint on the single particle action in that sector.
 We can now see how the underlying $\varphi$  action of eq.(\ref{act00})
 leads to the $\Phi$ action by inserting the constraint of eq.(\ref{norm000}) 
 onto the kinetic term of eq.(\ref{act00})
 and rearranging to obtain,
 \smbbo
\label{act1}
\backo
S=\!\!\int_{xy}\!\! \bl ZM^{2}|\varphi(y)\partial_x\varphi(x)|^2 
\nonumbo
\qquad \!\!\!
+{\frac{g^2M^2}{2}}\varphi^\dagger(y)\varphi(x)D_F(x-y)\varphi^\dagger(x)\varphi(y)\br
 \smebo
 and $S$ remains  dimensionless. With 
 the bilocal field of eq.(\ref{biloc})
 the bilocalized action becomes eq.(\ref{act0}).

We go to barycentric coordinates  $(X,r)$,
 \smbbo
\label{barycentric1}
\backo
X=\half(x+y),\qquad r=\half(x-y).
 \smebo
where $r^\mu=(r^0,\vvr)$, where $\vvr$ is the radius and $r^0$ is the relative time.

Hence we write,
 \smbbo
\backo
\Phi(x,y)=\Phi(X\!+\!r,X\!-\!r)\equiv
\Phi(X,r).
 \smebo
 Let  $S=S_K+S_P$ 
and can then rewrite 
the kinetic term, $S_K$,
using the derivative $\partial_x=\half(\partial_X\!+\!\partial_r)$, 
\smbbo
\label{act000}
\backo
S_K=
\frac{JM^4}{4} \!\! \int_{Xr}\!\! Z|(\partial_X\!+\!\partial_r)\Phi(X,r)|^2 
\nonumbo   \backo
=\frac{JM^4}{4} \!\!\int_{Xr}\!\bl Z|\partial_X\Phi|^2\!+\! Z|\partial_r\Phi|^2 
\nonumbo \qquad \qquad
\!+\! Z(\partial_X\Phi^\dagger\partial_r\Phi+h.c.)\br.
\smebo 
Note the Jacobian $J=16$,
\smbbo
J^{-1}=\left| \frac{\partial (X, r)}{\partial(x,y)} \right|=\bl \frac{1}{2}\br^4.
\smebo
Likewise, the potential term is,
 \smbbo
S_P=
\frac{JM^4}{2}\!\int_{Xr}\! {g^2}D_F(2r)|\Phi(X,r)|^2.
 \smebo
 
We will treat the latter term in eq.(\ref{act000}),
$
Z(\partial_X\Phi^\dagger\partial_r\Phi+h.c.),
$
as a constraint, 
with its contribution to the equation of motion,
\smbbo
\frac{\partial}{\partial X^\mu}\frac{\partial}{\partial r_\mu}\Phi=0.
\smebo
  We can redefine this term in the action as a Lagrange multiplier while
  preserving Lorentz invariance,
\smbbo
\label{LM}
\backo
\rightarrow \int_{Xr}\eta \bl
\frac{\partial\Phi^\dagger}{\partial X^\mu}\frac{\partial\Phi}{\partial r_\mu}+h.c.\br^2\;\;
 \makebox{hence,} \;\; \delta S/\delta \eta = 0,
 \smebo
which also enforces the constraint on a path integral in analogy to gauge fixing.
In the following we assume the constraint is present in the total action though not
written explicitly.
 
We therefore have the bilocal action, 
\smbbo
\label{23}
\backo\backo
S=
\frac{JM^4}{4}\!
\!\int_{Xr}\!\bl
Z|\partial_X\Phi|^2+ Z|\partial_r\Phi|^2 \!
\nonumbo \qquad 
+2g^2 D_F(2r)|\Phi(X,r)|^2\br.
\smebo
Following Yukawa, we assume we can factorize $\Phi$,
 \smbbo
\label{factor000}
\sqrt{J/4}\; \Phi(X,r)= \chi(X)\phi(r).
 \smebo
${\phi(r)}$ is the internal wave-function which we define to be dimensionless, $d=0$, 
while $\chi$ is an ordinary local field 
with mass dimension $d=1$.
$\chi(X)$ determines the center-of-mass motion of the composite state.
The action for the factorized field takes the form, 
\smbbo
\backo\backo
S=
{M^4}\!
\!\int_{Xr}\!\bl\!
Z|\partial_X\chi|^2|\phi^2|
\nonumbo
 + |\chi|^2( Z|\partial_r\phi|^2 \!
+2g^2 D_F(2r)|\phi(r)|^2\! ) \br.
\smebo
Matching the $U(1)$ current generated by 
$\chi\rightarrow e^{i\theta(X)}\chi$ to that of the underlying constituent theory
 we see that the normalization of the
world-scalar 4-integral is,\footnote{ The current is now generated by
$e^{i\theta(x)}\Phi=e^{i\theta((X+r)/2)}\Phi$. The matching is exact for the zero component (charge)
and implies eq.(\ref{chiact2}).  The currents will be
discussed in detail elsewhere.}
\smbbo
\label{chiact2}
1 =ZM^4 \! \int\! d^4r\; |\phi(r)|^2.
 \smebo
This then leads to a canonical normalization of $\chi(X)$.

We can then represent $S$ in terms of two ``nested'' actions. For the field $\chi$,
 \smbbo
\label{chiact}
\backo
S=
\!\int_{X}\!\bl |\partial_X\chi|^2-m^2|\chi|^2\br
\;\;\;\makebox{where,}\;\;\; m^2=- S_\phi,
 \smebo
and  $S_\phi$ is an action for the internal wave-function,
\smbbo
\label{act0p}
\backo\!\!\!
S_\phi=M^4\!\!\int_{r^0,\vvr}\!\bl\! Z|\partial_r\phi|^2\!
+{2g^2 }D_F(2r)|\phi|^2\!\br.
\smebo
Eq.(\ref{chiact}) then implies,
 \smbbo 
\partial^2_X\chi=-m^2\chi \;\;\;\makebox{hence},\;\;\chi\sim \exp(iP_\mu X^\mu).
 \smebo
$\chi(X)$ has free  plane wave solutions with $P^2=m^2$.

In the center of mass frame of the bound state we can choose $\chi$ to
have 4-momentum $P_\mu=(m,0,0,0)$
where we then have,
 \smbbo
\label{factorize}
\Phi(X,r)=\chi(X)\phi(r)\propto \exp(i m X^0)\phi(r).
 \smebo
$\phi(r)$ must then satisfy the Lagrange multiplier,
constraint 
\smbbo
P^\mu \frac{\partial}{\partial r^\mu} \phi(r) =0,
\smebo
and therefore becomes a {\em static function} of 
$r^\mu=(0,\vvr)$.

While we have specified $Z$ in eq.(\ref{chiact2}), we still have 
the option of normalizing the internal
wave-function
$\phi(\vvr)$. This can be conveniently 
normalized in the center of mass frame as,
 \smbbo
\label{norm01}
\backo\!\!\!\!\! M^3
\!\int \!\! d^3r\; |\phi(\vvr)|^2=1.
\smebo
Note that in eq.(\ref{norm01}) we have implicitly
defined the static internal wave-function $\phi(\vvr)$ to be 
dimensionless, $d=0$.

We see that the relative time now emerges in the 4-integral over $|\phi(r)|^2$
of eq.(\ref{chiact2}) together with eq.(\ref{norm01}),
 \smbbo
\label{norm0000}
\backo\backo
1=ZM^4\!\!\int\!\! d^4r|\phi(r)|^2=ZM^4\!\!\int \!\!dr^0 \!\!\!\int\!\! d^3r |\phi(\vvr)|^2
\nonumbo =ZMT,
 \smebo
 where $\int dr^0= \int dr^\mu P_\mu/M \equiv T$.
Then from eq.(\ref{norm0000}) we have,
\smbbo
\label{norm02}
TZ=M^{-1}.
 \smebo
With static $\phi(r)\rightarrow \phi(\vvr)$ the internal  action of eq.(\ref{act0p}) becomes,
\smbbo
\label{SKG0}
\backo
S_\phi=M^4\!\!\int_{r^0,\vvr}\!\bl\!-Z|\partial_{\vvr}\phi(\vvr)|^2\!
+{2g^2 }D_F(2r)|\phi(\vvr)|^2\!\br,
\smebo
where $|\partial_{\vvr}\phi(\vvr)|^2=\partial_{\vvr}\phi^\dagger\cdot\partial_{\vvr}\phi$.
Note  $\partial_{\vvr}\phi$ is spacelike,
and the arguments of the constrained $\phi(\vvr)$ are now 3-vectors,
however $D_F(2r^\mu)$ still depends upon the 4-vector $r^\mu$.

There remains the integral over relative time $r^0$ in the action.
For the potential, we have by residues,
\smbbo
\label{42}
\backo\!\!\!\!\!\!\!  V(r)=-2\!\!\int \!\!dr^0 D_F(2r)
 =-\!\int \!\!\frac{e^{2i\vvq\cdot\vvr}}{{\vvq}^2+M^2} \frac{d^3q}{(2\pi)^3}
 \nonumbo
 =-\!\frac{e^{-2M|\vvr|}}{8\pi |\vvr|},
\smebo 
and the potential term becomes the static Yukawa potential,
\smbbo
\backo
S_P=-M^3\!\!\int_{\vvr}\!
{g^2 }MV(\vvr)|\phi(\vvr)|^2,\;\;\;
\smebo
The $\phi(\vvr)$ kinetic term in eq.(\ref{SKG0}) becomes,
\smbbo
\backo
S_K=-M^4\!\!\int_{r^0,\vvr}\!\!Z|\partial_{\vvr}\phi(\vvr)|^2\!
=-M^4ZT\!\! \int_{\vvr}\! |\partial_{\vvr}\phi(\vvr)|^2\!
\nonumbo
=-M^3\!\!\int_{\vvr}\! |\partial_{\vvr}\phi(\vvr)|^2\!,
\smebo
where we use eq.(\ref{norm02}).
The action $S_\phi$ thus becomes, 
\smbbo
\label{act0ppp}
\backo\!\!\!
m^2=
-S_\phi= M^{3}\int_{\vvr} \bl |\partial_{\vvr}\phi|^2
+{g}^2MV(|\vvr|)|\phi|^2 \br.
\smebo
Note $S_\phi$ has dimension $d=2$, as it must for $m^2$.
We thus see, as previously mentioned, that the combination $ZT$ occurs in the theory, and 
the relative time has disappeared into normalization constraints,
eqs.(\ref{chiact2}, \ref{norm01}).

The extremalization of $S_\phi$ leads to the  Schr\"odinger-Klein-Gordon (SKG) equation,
\smbbo
\label{acteqn1}
\backo
-\nabla_{\vvr}^2\phi(r) -{g^2M}\frac{e^{-2M|\vvr|}}{8\pi|\vvr|}\phi(r)= m^2{\phi}(r),
\smebo
where, for spherical symmetry in a ground state,
\smbbo
\nabla_r^2=\partial_r^2+\frac{2}{r}\partial_r.
\smebo
We see that the induced mass$^2$ of the bound state, $m^2$,
is the eigenvalue of the SKG equation.
We can compare this to a non-relativistic Schr\"odinger equation (NRSE),
 \smbbo
\label{acteqn1}
\backo
-\frac{1}{2M}\nabla_r^2{\phi}(r) - {g^2}\frac{e^{-2Mr}}{16\pi r}{\phi}(r)= E{\phi}(r).
 \smebo
 In the next section we will obtain a similar results for
 a bound state of chiral fermions and use the known results for the Yukawa potential
 in the NRSE to obtain the critical coupling.
 The negative eigenvalue of $E$ in the NRSE, which signals binding, presently implies
 a vacuum instability.
 
Integrating by parts we then have from eq.(\ref{act0ppp}),
\smbbo
\label{act0pppp}
\backo\!\!\!
m^2 =M^{3}\int_{\vvr} \bl \phi^\dagger (-\nabla^2_{\vvr}\phi
+{g^2M}V(r)\phi(\vvr))\br.
\smebo
Note consistency, using eq.(\ref{acteqn1}),
and the normalization of the dimensionless field $\phi$
 of eq.(\ref{norm01}).

More generally, 
by promoting $\chi$ to a $(1+3)$ time dependent field while maintaining a static $\phi$
we have the full joint action:
 \smbbo
\backo\;
S=\!
M^3\!\!\int_{X\vvr} \!\bl \!|\phi|^2\!\left|\frac{\partial\chi}{\partial X}\right|^2\!
\!\!-\!|\chi|^2\bl |\partial_{\vvr}\phi|^2\!
+\!g^2M V(\vvr)|\phi|^2\br\br .
\nonumbo
 \smebo

In summary, we have constructed, by ``bilocalization''
of a local  field theory, a bilocal field description
$\Phi(x,y)$ for the dynamics of binding
a pair of particles.  The dynamics
implies that, in barycentric coordinates, $\Phi(x,y)\sim \Phi(X,r)\sim \chi(X)\phi(\vvr)$,
where the internal wave-function, $\phi(\vvr)$, is a static function of $\vvr$
and satisfies an SKG equation with eigenvalue $m^2$, which determines the squared-mass of
a bound state.  This illustrates the removal of relative time in an action formalism, which
is usually framed in the context of Dirac Hamiltonian constraints \cite{Dirac}.

\subsection{The Coloron Model $\label{3}$}

  The point-like NJL model can be viewed as the limit of a physical theory
with a bilocal interaction.
An example that
 motivates the origin of the NJL is the   ``coloron model,'' deployed
 in Sections III and IV
to describe chiral constituent quark models 
\cite{Bijnens}\cite{BHHQ}\cite{Topcolor}\cite{ NSD}\cite{Simmons}.

We assume chiral fermions, each with $N_c$ ``colors'' labeled by $(a,b,...)$
with the local Dirac action,
 \smbbo
 \label{fermikinetic}
\backo
S_F= \int_x\bl i\bar{\psi}^a_L(x)\slash{\partial}\psi_{aL}(x)+ i\bar{\psi}^a_R(x)\slash{\partial}\psi_{aR}(x)\br,
 \smebo
The single coloron exchange interaction   
 then takes a bilocal current-current form:
\smbbo
\label{TC0}
 \backo\;
S_C=-g^2\!\!\!\int_{xy} \!\!
 \bar{\psi}_{L}(x) \gamma_\mu T^A \psi_L(x)
 \nonumbo \qquad \times \;D^{\mu\nu}(x-y)\;
\bar{\psi}_{R}(y) \gamma_\nu T^A \psi_{R}(y),
\smebo
where $T^A$ are generators of $SU(N_c)$. 
The coloron propagator  in a Feynman gauge yields:
\bea
\label{propagator}
D_{\mu\nu}(x-y)=\int\frac{-ig_{\mu\nu}}{q^2-M^2}e^{iq(x-y)}\frac{d^4q}{(2\pi)^4}.
\eea
A Fierz rearrangement of the interaction to leading order in $1/N_c$ 
leads to an attractive potential \cite{Topcolor}:
\smbbo
\label{coloronexchange}
\backo\backo
S_C=g^2\!\!\int_{xy}\!\! \; \bar{\psi}^a_L(x)\psi_{aR}(y)\; D_F(x-y)\;\bar{\psi}^b_R(y)\psi_{bL}(x),
\smebo
where $D_F$ is defined in eq.(\ref{DF}).
Note that if we suppress the $q^2$ term in the denominator
of eq.(\ref{propagator}), 
 \smbbo
\label{propagator2}
{D}_F(x-y)\rightarrow  \frac{1}{M^2}\delta^4(x-y),
 \smebo
and  we  immediately recover the point-like NJL model interaction.

Consider spin-$0$ fermion pairs of a given color $[\bar{a}b]$
$ \bar{\psi}^a_R(x)\psi_{bL}(y)\!$.
We 
will have free fermionic scattering states, $\sim\; :\! \bar{\psi}^a_R(x)\psi_{bL}(y)\! :$,
coexisting in the action with bound states $\Phi(x,y)$,
 \smbbo
\label{act2}
\backo\!\!\! \bar{\psi}^a_R(x)\psi_{bL}(y)\!
\rightarrow \;:\! \bar{\psi}^a_R(x)\psi_{bL}(y) \!:+ \; M^2\Phi^a_{\;b}(x,y).\;\;
 \smebo 
 This is analogous to the shift done to introduce the auxiliary field
 in the NJL model in eq.(\ref{NJL2}). The main differences with the NJL model are that presently
 $\Phi^a_{\;b}(x,y)$ is bilocal and it has a bare two body kinetic term.
 The normal ordering $:...:$ signifies that we have subtracted
 the bound state from the product.

We see that
$\Phi^a_b(X,r)$ is an $N_c\times N_c$ complex matrix that transforms as a product
of $SU(N_c)$ representations,  $\bar{N}_c\times  {N}_c$,
and therefore decomposes into a singlet plus an adjoint representation of $SU(N_c)$.
We write $\Phi^a_b$
it as a matrix  $\widetilde\Phi$  by introducing the $N_c^2-1$ adjoint matrices,
${T}^A$, where $\Tr ({T}^A{T}^B)=\half\delta^{AB}$.
The unit matrix is ${T}^0\equiv\;$diag$(1,1,1,..)/\sqrt{2N_c}$, 
and $\Tr({T}^0{})^2 = 1/2$, hence we have,
\smbbo
\widetilde{\Phi} =\sqrt{2}({T}^0\Phi^0 + \sum_{A}{T}^A\Phi^A).
\smebo
The $\sqrt{2}$ is present because $\Phi^0$ and $\Phi^A$ form complex representations  
since they also represent the $U(1)_L\times U(1)_R$
chiral symmetry.

For the bilocal fields, 
we have a  bosonic kinetic  term  which,
for the singlet representations, takes the form,
\smbbo
\label{act1}
\backo
S_K= \frac{{J}ZM^4}{2}
\nonumbo \times \int_{Xr}\! \bl|\partial_X\Phi^0|^2+|\partial_r\Phi^0|^2
\!+\! \eta|\partial_X\Phi^0{}^\dagger\partial_r\Phi^0|^2
\br .
\smebo
We assume the constraint in the barycentric frame, and integrate out
relative time with $ZMT=1$,
\smbbo
\label{act1}
\backo\backo
S_K= (J/2)\!\!\int_{X}\int'_{\vvr}\! \bl|\partial_X\Phi^0(X,\vvr)|^2
-|\partial_{\vvr}\Phi^0(X,\vvr)|^2
\br
\smebo
(where $\int'_{\vvr} =M^3\int d^3r$).

The full interaction of eq.(\ref{coloronexchange}) thus becomes,\small
\smbbo 
\label{interaction00}
S_C
\rightarrow
g^2\!\!\int_{xy}\!\! :\!\bar{\psi}^a_L(x)\psi_{aR}(y)\!: D_F(x-y) :\!\bar{\psi}^b_R(y)\psi_{bL}(x)\!:
\nonumbo 
+g^2JM^2\sqrt{N_c}\!\!\int_{X,r}\!\!\!\!\!:\!\bar{\psi}^a_L(X\!-\!r)\psi_{aR}(X\!+\!r)\!:\!\! D_F(2r)\;\Phi^0{+h.c.}
\nonumbo
+\;g^2JM^4N_c\!\!\int_{X,r}\!\!\!\! \; \Phi^0{}^\dagger(X,r)\; D_F(2r)\;\Phi^0(X,r)
\smebo
\normalsize
where,
 \smbbo
{D}_F(2r)= -\!\int\frac{1}{(q^2-M^2)}e^{2iq_\mu r^\mu}\frac{d^4q}{(2\pi)^4}.
 \smebo
The leading term $S_C$ of eq.(\ref{interaction00}) is just a free 4-fermion
scattering state interaction 
and has the structure of the NJL interaction in the limit of eq.(\ref{propagator2}).
This identifies $g^2$ as the NJL coupling constant.
This is best treated separately by the local interaction of eq.(\ref{coloronexchange}). 
We therefore omit this term in
the discussion of the bound states.

The second term $\sim \Tr(\psi^\dagger\psi)\Phi^0+h.c.$ in eq.(\ref{interaction00}) 
determines the Yukawa interaction between
the bound state $\Phi^0$ and the free fermion scattering states.
We will treat this below.

Note that the third term in eq.(\ref{interaction00})  is the binding interaction and it
involves only the singlet, $\Tr\widetilde\Phi = \sqrt{N_c}\Phi^0$. 
The adjoint representation 
 $\Phi^A$ are decoupled from the interaction
 and remain as unbound, two body massless scattering states. 
Moreover, the singlet field $\Phi^0$ therefore has an enhanced
 interaction by a factor of $N_c$.
This is analogous to a BCS superconductor, where the $N_c$ color pairs
are analogues of $N$ Cooper pairs and the weak 4-fermion Fr\"ohlich Hamiltonian interaction is enhanced 
by a factor of $N_{Cooper}$ \cite{Cooper}.
The color factor enhancement also occurs in the NJL model, but at loop level.
Here we see that the color enhancement is occurring in the semi-classical (no loop) coloron 
theory by this coherent mechanism.  To our knowledge this semi-classical enhancement
of the underlying coupling strength in a coloron model, or its
approximation to QCD chiral dynamics, has not been previously noted.

Factorizing $\Phi^0$,
 \smbbo
\label{factor000}
\sqrt{J/2}\; \Phi^0(X,r)= \chi(X)\phi(r),
 \smebo
the kinetic term action becomes identical to the bosonic case,
\smbbo
\label{act1}
\backo\backo
S_K= \!\!\int_{X} \bl|\partial_X\chi(X)|^2-|\chi(X)|^2\int'_{\vvr}\!|\partial_{\vvr}\phi(\vvr)|^2
\br,
\smebo
with,
\smbbo
\int'_{\vvr}|\phi(\vvr)|^2=1.
\smebo
The interaction term
can then be written from eq.(\ref{42}) as,
\smbbo
\label{interaction0011}
\backo
S_C
\rightarrow
g^2JM^4N_c\!\!\int_{X,r}\!\!\!\! \; \Phi^0{}^\dagger(X,r)\; D_F(2r)\;\Phi^0(X,r)
\nonumbo
=
g^2N_c\!\int_{X} |\chi(X)|^2 \int'_{\vvr}|\phi(\vvr)|^2 M\frac{e^{-2M|\vvr|}}{8\pi |\vvr|}.
\smebo
Hence, the removal of relative time 
is identical procedure as in the bosonic model,
and we obtain to the same action, $S=S_K+S_C$.

The extremalization of $\phi$ then leads to the 
SKG equation,
\smbbo
\label{acteqn2}
\backo
-\nabla_{\vvr}^2\phi(\vvr) -{g^2N_c M}\frac{e^{-2M|\vvr|}}{8\pi|\vvr|}\phi(\vvr)= m^2{\phi}(\vvr).
\smebo

\subsection{Classical Criticality of the Coloron Model }

The coloron model  furnishes a  direct UV
completion of the NJL model.  However, in the coloron model we do not
need to invoke large-$N_c$ quantum loops to have a critical theory.
Rather, it leads to an SKG potential of the Yukawa form
which has a {\em classical critical coupling}, $g_c$.
For $g<g_c$ the theory is subcritical and produces
resonant bound states that decay into chiral fermions.
For $g>g_c$ the theory produces a tachyonic bound state which implies
a chiral instability and $\Phi$ must develop a VEV. This
requires stabilization by, e.g., quartic interactions and a sombrero potential.
All of this is treated bosonically in our present formalism.

The criticality of the Yukawa potential in the nonrelativistic
Schr\"odinger equation is discussed in the literature in the context of ``screening.''
The nonrelativistic Schr\"odinger equation $r=|\vvr|$ is:
 \smbbo
-\nabla^2\psi - 2m\alpha\frac{e^{-\mu r}}{r}\psi=2mE
 \smebo
and criticality (eigenvalue $E=0$) occurs for $\mu=\mu_c$ where a numerical analysis yields,
\cite{Edwards},
 \smbbo
\mu_c= 1.19\;\alpha m.
 \smebo
For us the spherical SKG equation is now  eq.(\ref{acteqn2}).  
Comparing,  gives us a critical value of the coupling constant,
when $\mu_c\rightarrow 2M$, $m\rightarrow M/2$ and $\alpha\rightarrow g^2N_c/8\pi$, then:
 \smbbo
 \label{critc1}
\backo\!\!\!
2M=(1.19)\bl\!\frac{M}{2}\!\br\bl\!\frac{g^2N_c}{8\pi}\!\br, \;\;\makebox{hence:}\;\;
{g^2}/{4\pi}=
6.72/N_c.\nonumbo
 \smebo
We can compare the NJL critical value of eq.(\ref{NJL3}), 
\smbbo
\label{critc2}
g_{cNJL}^2/4\pi = 2\pi/N_c=6.28/N_c.
\smebo
Hence, the NJL quantum criticality is a comparable effect, with a remarkably
similar numerical value for the critical coupling.

Note that we can rewrite eq.(\ref{acteqn2}) with dimensionless coordinates, $\vec{u}=M \vvr$,
\smbbo
\label{crit4}
M^2\bl-\nabla_u^2\phi(u) -{g^2N_c}\frac{e^{-2u}}{8\pi u}\br\phi(u)= 0,
 \smebo
and then $M^2$ only appears as an overall scale factor.   Hence we see that critical coupling is
determined by eq.(\ref{crit4}), and the scale $M$ cancels out at criticality. 
The mass  scale $M$ is dictated
in the exponential $e^{-2Mr}$ of the Yukawa potential.  
However, in general, we can start with the dimensionless form of
the SKG equation, as in eq.(\ref{crit4}),
and infer the scale $M$ by matching to any desired potential. 

However, it is important to realize that the NJL model involves
the Yukawa coupling, $g_{NJL}$ while the present criticality involves the 
coloron
coupling constant.
The Yukawa coupling is emergent in the coloron model, and we need to compute it.

The second term in eq.(\ref{interaction00}) is the induced the Yukawa interaction $S_Y$,
and can be written as:
\smbbo
\label{Yukawa000}
S_Y=g^2(\sqrt{{2JN_c}})M^2\times
\nonumbo
\int_{Xr}\!\bar{\psi}^a_L(X-r)\psi_{aR}(X+r) D_F(2r) \chi(X) \phi(\vvr){+h.c.}
\smebo
This is the effective Yukawa interaction between
the bound state $\Phi^0$ and the free fermion scattering states.

We can't simply integrate out the relative time
here. 
However, we can first connect this to the point-like limit by suppressing
 the $q^2$ term in the denominator of $D(2r)$ with $z\rightarrow 0$ in:
 \smbbo
 \backo
 {D}_F(2r)\rightarrow \!\!\int\!\!\frac{1}{M^2}e^{2iq_\mu r^\mu}\frac{d^4q}{(2\pi)^4}
 \rightarrow \frac{1}{JM^2}\delta^4(r),
 \smebo
 where $\delta^4(2r)=J^{-1}\delta^4(r)$,
 hence with $J^{-1}=1/16$,
 \smbbo
 \label{74}
 \backo
 S_Y=g^2(\sqrt{{N_c}/{8}})
\int_{X}\!\bar{\psi}^a_L(X)\psi_{aR}(X) \chi(X) \phi(0)+h.c.
\nonumbo
 \smebo
 This gives the Yukawa coupling,
 \smbbo
 \label{gY}
 g_Y=g^2(\sqrt{{N_c}/{8}})\phi(0).
 \smebo
 The wave-function at the origin, $\phi(0)$ in the NJL limit
 is somewhat undefined.  However, if we consider a spherical cavity
 of radius $R$ where $MR=\pi/2$, with a confined,  dimensionless, $\phi(r)$,
 then $\phi(0)$ is obtained:
\smbbo
\phi(0)= \frac{1}{\pi}.
\smebo
Plugging this into the expression for $g_Y$ in eq.(\ref{gY}) gives,
\smbbo
\label{77}
g_{Y}  =g^2\sqrt{{N_c}/{8\pi^2}}=\frac{g^2}{g_{cNJL}},
\smebo
where $g^2_{cNJL}=(8\pi^2/N_c)$ is the critical coupling of the
NJL model, as seen in eq.(\ref{NJL3}).

Hence, if the coloron coupling constant,  $g^2$,  is critical, as in eqs.(\ref{critc1},\ref{critc2}).
then we have seen in that $g^2\approx g^2_{cNJL}$,
and  the induced Yukawa coupling from eq.(\ref{77}) is then $g_Y\approx g_{cNJL}$.
The coloron model is then consistent with the NJL model
in the point-like limit where the NJL model coupling is the Yukawa coupling, as seen in eq.(\ref{NJL2}).   
However, the induced Yukawa coupling
in the bound state, $g_Y$, may be (significantly?) different than the coloron coupling $g$
in realistic extended $\vvr$ models.

For subcritical
coupling there are resonance solutions with positive $m^2$ that have large distance tails of 
external incoming and outgoing radiation, representing a steady state
of resonant production and decay. 

With super-critical coupling, $g>g_c$, the bilocal field $\Phi(X,r)$ has a negative squared mass
eigenvalue (tachyonic), with a well-defined localized wave-function.  In the region external to the
potential (forbidden zone)
the field is exponentially damped. At exact criticality with $g=g_c-\epsilon$ there is a $1/r$
(quasi-radiative) tail that switches to exponential damping for $g=g_c+\epsilon$.
The supercritical solutions are localized and normalizable over the
entire space $\vvr$, but with $m^2<0$ they lead to exponential
runaway in time of the field $\chi(X^0)$, and must be stabilized, typically
with a $|\Phi|^4$ interaction.

We then treat the supercritical case as resulting in spontaneous symmetry breaking. 
In the point-like  limit, $\Phi(X)\sim \Phi(X,0)$, 
the theory has the ``sombrero potential'', 
  \smbbo
 V(\Phi) = -|M^2\Phi^2|+\frac{\lambda}{2}|\Phi|^4.
  \smebo
The point-like field develops a VEV, $\langle \Phi \rangle =|M|/\sqrt{\lambda}$.
In this way the bound state theory will drive 
the usual chiral symmetry breaking
from the underlying dynamics of a potential induced by new physics. 

The simplest sombrero potential can be modeled
as,
 \smbbo
\backo\;
S=\!\!\int_{X\vvr}' \!\bl \!|\phi|^2\!\left|\frac{\partial\chi}{\partial X}\right|^2\!
\!\!-\!|\chi|^2( |\nabla_r\phi|^2\!
+\!g^2N_cM V(r)|\phi(r)|^2) 
\nonumbo
\qquad\qquad
-|\chi|^4\frac{\hat\lambda}{2}|\phi(\vvr)|^4\br.
 \smebo
In the case of a perturbatively small $\lambda$
we expect the eigensolution of $\phi$
to be essentially unaffected,
\smbbo
\int'_r \!\! \bl \!|\nabla_{\vvr}\phi|^2\!
+\!{g^2N_c\!M}V(\vvr)|\phi(\vvr)|^2\!\br \approx m^2.
\smebo
The effective quartic coupling is then further renormalized by
the internal wave-function,
\smbbo
\frac{\hat \lambda}{2}|\chi|^4\int'_r |\phi(\vvr)|^4=|\chi|^4\frac{\widetilde\lambda}{2}.
\smebo
In this case we
 see that $\chi$ develops a VEV in the usual way:
 \smbbo
\langle |\chi|^2 \rangle = |m^2|/\widetilde\lambda =v^2.
 \smebo
This is consequence of $\phi(\vvr)$ remaining localized in its potential

The external scattering state fermions, $\psi^a(X)$, will then acquire mass 
through the emergent
Yukawa interaction described in the previous section, $\sim g_Y\langle |\chi| \rangle.$
 For general $\widetilde\lambda$ and possibly large
 (as in a nonlinear sigma model),
 the situation is potentially more complicated.
 While perturbative 
 solutions maintain locality in $\phi(\vvr)$, it is unclear what solutions exist 
  for non-perturbative $\lambda$.
 

\newpage

\section{A Personal recollection of Yoichiro Nambu}

In the 1990's Nambu had taken a considerable interest in the idea of top
quark condensation, and we
became friends.     
During this era, and
on several occasions, he would surprisingly appear at my office door at Fermilab. 
I sensed he enjoyed an occasional sojourn away from the city, and perhaps
the open space of cornfields and the lab had some appeal. On these visits
we would have discussions on far ranging topics, including historic
Japanese baseball games, bridge building, and his basement experiments
(Bruce Winstein told me years earlier that Nambu had a sophisticated optics lab in his basement
and a penchant for experiment).

Nambu's thinking about physics was subtle and connections
that interested him were often seemingly remote to others. Some examples were
smoke rings and strings,  Bode's law of planetary orbits and naturalness, and
that he thought there was a hidden supersymmetry
in the NJL model (which inspired me to study this question and
led to an ``approximate SUSY'' in \cite{super} which may be the leading terms
in a power series in $\Lambda^{-2}$, but I consider the question
still open).  Nambu considered that the heavy top quark may be special and 
that it may play a significant role 
in electroweak physics, which is certainly a view I shared and still have to this day.

In 1992 I attended a meeting in Hiroshima \cite{Hiroshima} on electroweak symmetry breaking and gave
a talk on top quark condensation.  My talk was a long one, and was last of the morning
session, followed by a bus that would take the speakers
to  a fine local restaurant for lunch.
At the conclusion of my talk, however, I was surrounded by a number of
students and post-docs with many questions about technical details.
Mainly,  there was extant confusion as to whether a particular bubble-sum calculation
was better than another.  I was mainly trying to dispel this,
arguing that the renormalization group is
the unique and far superior way to view the dynamics of NJL.
The discussion ran on for almost a half an hour and
I missed the bus.

I finally exited the large conference center onto a plaza and studied the surrounding streets
in search of a convenient restaurant for lunch (and a beer).
I couldn't spot one, and I expected I'd have to
walk some distance from the plaza, when suddenly someone tapped me on the shoulder...it was Nambu!

Nambu said, ``I am taking you to lunch to a special place for the Japanese delicacy, called
`okonomyaki,' which is particularly good here."  
So we hiked several blocks to a subway, which took us to a transfer where
we boarded a street tram.  I was struck by the tram, which was old, dark green, and wooden and
resembled photos of those I had seen of Hiroshima after the war. The tram progressed to the coast and
 to the harbor, which was modern and active with large container ships and
 a buzz of loading--unloading operations
 underway.
 Here we came upon a  diner, 
 crowded with workers, and we were seated at a small table. Nambu
 did the ordering, and soon our plates of okonomiyaki arrived with 
 Asahi beers.
 
 The okonomiyaki was for me a bit problematic, though possibly it could
 become an acquired taste. 
 It seemed to be a kind of pizza with the parts of the fish 
 partially grilled on top, some more or less so, 
 and featured {\em all} of the parts of fish (including innards) as part of the recipe.
 I mustered a slight smile and nod to Nambu who could clearly tell my relishing 
 of the delicacy was somewhat muted. So I struggled for some conversational gratuity,
 and I asked him about his early days in physics. 
 
 He talked about the end of the war and his first post-doc
 at the Institute for Advanced Study.  
 Oppenheimer was in charge and Albert Einstein was in residence.  
 Nambu had nearly completed
 his post-doctoral appointment and had befriended Einstein's secretary.
 She was interested in Japanese culture, and practiced a little 
 of the language with him
 (I am not sure if this was Helen Dukas or someone else).
 
 As the end of his appointment approached,
 Nambu asked if it would be possible to meet Professor Einstein?
 His secretary said she could arrange it, but it was very important that Oppenheimer
 not find out. Oppenheimer had put out an edict that ``no-one shall bother
 Prof. Einstein,'' and Oppenheimer could be quite imperious in his management style.
 So, an appointment was scheduled late in the day,
 5 PM or so, when Einstein would still be working quietly in his office.
 
 On the appointed day Nambu arrived at Einstein's office door and gently knocked.
 From within came the distinctive voice, ``come in.''   Nambu entered and bowed in Japanese style
 to the professor who was seated at his desk, pipe in hand.  After introduction,
 Einstein said, ``So what are you working on?"\footnote{I later heard from Leon Lederman
 of his two encounters with Einstein, that this was the standard greeting. You could
 then expect about a minute to explain what you were doing and Einstein would then
 reply. In Leon's case, he described the K-mesons on one occasion, and 
 neutrinos on another. Each time Einstein's reply was
 "achh, so many particles.  If I wanted to study so many species I would have been a
 botanist!"}
 
 Nambu had been working on collective phenomena in condensed matter physics, superconductivity,
 Ising models, etc., topics which clearly influenced his later
 thinking about symmetry breaking in particle physics.  I didn't quite catch specifically what
 problem he described to Einstein 
 but after a minute Einstein reacted.
 
 The Moon was rising in the early evening and visible through Einstein's office window.
Einstein was tamping tobacco into his pipe with a piece of chalk.
 He then pointed the chalk toward the Moon and said, ``Tell me...do you think that does not 
 simultaneously have a position and a momentum?''  There was momentary silence.  
 
 Einstein then launched into an
 animated diatribe against the quantum theory. 
 He evidently became rather exercised in this, and Nambu
 was terrified that he had so upset the greatest scientist in human history. 
 After a few minutes Nambu stood up, bowed, humbly backed away, thanking the 
 professor, and made a hasty retreat.
 
 Nambu was mortified by this experience, but
 to add to his misery Oppenheimer found out about the encounter and was furious
 and summoned Nambu to his office for a stern admonishment.
 Nambu became depressed  in his last few weeks at the IAS, and packed his bags and prepared
 to return, jobless, to Japan.  Literally, however, the last time he went to
 collect his mail at the IAS, there was a letter... from The University of Chicago
 and it was an offer for him to join the faculty.
 
 We finished our okonomyaki and made the reverse trip
 back to the conference center.  
 I wondered then, and have ruminated ever since, about the strange
 irony of this event: 
 
 Hiroshima: I had lunch with the great physicist, Yoichiro Nambu, who 
 explained the broken chiral symmetry of strong interaction mass and the special properties 
 of nearly massless pi-mesons as a general phenomenon; 
  who had a remarkable encounter with Albert Einstein,
 who created modern physics, gave us $E=mc^2$, General Relativity,
 and had even introduced the idea of photons as quanta (though
 rejected the formulation of the quantum theory);
  only to infuriate J. Robert Oppenheimer, 
 who managed the construction of the bomb that destroyed: $\;\;$ Hiroshima.
 
Years later I was fortunate to attend the Nobel luncheon for Nambu at The University of Chicago
in 2008.
Nambu was unable to travel to Stockholm due to his wife's illness, and the Swedish ambassador
arranged to present the medal to him in Chicago later that afternoon.
The luncheon  delightfully turned out to be a small gathering of about a dozen  people,
including Nambu's son, John.

 At the luncheon the conversation lulled. 
 Nambu was happy to
 achieve this award, but at a deeper level there was a visible sadness at the prospect
 of his wife's passing.  So I asked him to tell the Einstein story
 to the luncheon table. He did so, and I felt that momentarily he may have been lifted by the diversion.
 
 The Nobel luncheon was my last encounter with Nambu, who  retired and returned 
 to Japan, and passed away in 2015.  Ironically, Nambu had also won the Oppenheimer
 Memorial Prize in 1977.

 \newpage

\section*{Acknowledgments}
I thank
the  Fermi Research Alliance, LLC under Contract No.~DE-AC02-07CH11359 
with the U.S.~Department of Energy, 
Office of Science, Office of High Energy Physics, and the University
of Wisconsin, Madison.

\end{document}